\documentclass{aa}

\usepackage{graphicx}
\usepackage[graphicx]{realboxes}
\usepackage{txfonts}
\begin{document}

   \title{Formation of complex molecules in translucent clouds: Acetaldehyde, vinyl alcohol, ketene, and ethanol via “nonenergetic” processing of C$_2$H$_2$ ice
   	}
\titlerunning{“nonenergetic” processing on C$_2$H$_2$ ice}
   \author{K.-J. Chuang\inst{1,2}, G. Fedoseev\inst{3}\thanks{current address:Laboratory for Astrophysics, Leiden Observatory, Leiden University, P.O. Box 9513, NL-2300 RA Leiden, the Netherlands}, D. Qasim\inst{1}, S. Ioppolo\inst{4}, C. Jäger\inst{2}, Th. Henning\inst{5},  M. E. Palumbo\inst{3}, E.F. van Dishoeck\inst{6}, H. Linnartz\inst{1}
    }
\authorrunning{Chuang et al.}
   \institute{Laboratory for Astrophysics, Leiden Observatory, Leiden University, P.O. Box 9513, NL-2300 RA Leiden, the Netherlands
              \and
              Laboratory Astrophysics Group of the Max Planck Institute for Astronomy at the Friedrich Schiller University Jena, Institute of Solid State Physics, Helmholtzweg 3, D-07743 Jena, Germany
              \and
              INAF – Osservatorio Astrofisico di Catania, via Santa Sofia 78, 95123 Catania, Italy
              \and
              School of Electronic Engineering and Computer Science, Queen Mary University of London, 
              Mile End Road, London E1 4NS, UK
              \and
              Max Planck Institute for Astronomy, Königstuhl 17, 69117 Heidelberg, Germany
              \and
              Leiden Observatory, Leiden University, P.O. Box 9513, NL-2300 RA Leiden, the Netherlands\\
              \email{chuang@mpia.de}
              }

  \date{}

 
  \abstract
   {Complex organic molecules (COMs) have been identified toward high- and low-mass protostars as well as molecular clouds, suggesting that these interstellar species originate from the early stage(s) of starformation. The reaction pathways resulting in COMs described by the formula C$_2$H$_\text{n}$O, such as acetaldehyde (CH$_3$CHO), vinyl alcohol (CH$_2$CHOH), ketene (CH$_2$CO), and ethanol (CH$_3$CH$_2$OH), are still under debate. Several of these species have been detected in both translucent and dense clouds, where chemical processes are dominated by (ground-state) atom and radical surface reactions. Therefore, efficient formation pathways are needed to account for their appearance well before the so-called catastrophic CO freeze-out stage starts.}
   {In this work, we investigate the laboratory possible solid-state reactions that involve simple hydrocarbons and OH-radicals along with H$_2$O ice under translucent cloud conditions (1$\leq$A$_V$$\leq$5 and \textit{n}$_\text{H}$$\sim$10$^3$ cm$^{-3}$). We focus on the interactions of C$_2$H$_2$ with H-atoms and OH-radicals, which are produced along the H$_2$O formation sequence on grain surfaces at 10 K.}
   {Ultra-high vacuum (UHV) experiments were performed to study the surface chemistry observed during C$_2$H$_2$ + O$_2$ + H codeposition, where O$_2$ was used for the in-situ generation of OH-radicals. These C$_2$H$_2$ experiments were extended by a set of similar experiments involving acetaldehyde (CH$_3$CHO) – an abundant product of C$_2$H$_2$ + O$_2$ + H codeposition. Reflection absorption infrared spectroscopy (RAIRS) was applied to in situ monitor the initial and newly formed species. After that, a temperature-programmed desorption experiment combined with a Quadrupole mass spectrometer (TPD-QMS) was used as a complementary analytical tool. The IR and QMS spectral assignments were further confirmed in isotope labeling experiments using $^{18}$O$_2$.}
   {The investigated 10 K surface chemistry of C$_2$H$_2$ with H-atoms and OH-radicals not only results in semi and fully saturated hydrocarbons, such as ethylene (C$_2$H$_4$) and ethane (C$_2$H$_6$), but it also leads to the formation of COMs, such as vinyl alcohol, acetaldehyde, ketene, ethanol, and possibly acetic acid. It is concluded that OH-radical addition reactions to C$_2$H$_2$, acting as a molecular backbone, followed by isomerization (i.e., keto-enol tautomerization) via an intermolecular pathway and successive hydrogenation provides a so far experimentally unreported solid-state route for the formation of these species without the need of energetic input. The kinetics of acetaldehyde reacting with impacting H-atoms leading to ketene and ethanol is found to have a preference for the saturated product. The astronomical relevance of the reaction network introduced here is discussed.}
   {}

   \keywords{astrochemistry – methods: laboratory: solid state – infrared: ISM – ISM: atoms – ISM: molecules – molecular
   	Processes}

\maketitle{}
\section{Introduction}   
Astronomical infrared observations in star-forming regions have revealed ice features of a number of abundant species, such as H$_2$O, CO$_2$, CO, CH$_3$OH, NH$_3$, and CH$_4$ (see review \citealt{Boogert2015} and references therein). The origin of these interstellar ices has been intensively studied, both observationally and in the laboratory, showing that most interstellar ices comprise of two separate ice layers, namely a H$_2$O-rich and a CO-rich layer on top that follow two sequential accretion stages \citep{Tielens1991,Gibb2004,Pontoppidan2006,Oberg2011b,Mathews2013,Boogert2015}. Recent laboratory studies focusing on grain surface chemistry have shown that CO hydrogenation at temperatures as low as $\sim$13 K efficiently leads to the formation of H$_2$CO and CH$_3$OH \citep{Watanabe2003,Fuchs2009}. Moreover, it offers a nonenergetic (i.e., in the absence of processing by high energetic particles) pathway toward O-bearing species containing multiple carbon atoms (chemical formula: C$_2$H$_\text{n}$O$_2$ and C$_3$H$_\text{m}$O$_3$), such as sugar-like molecules (e.g., glycolaldehyde and glyceraldehyde) and sugar alcohols (e.g., ethylene glycol and glycerol) \citep{Fedoseev2015b,Fedoseev2017,Butscher2015,Butscher2017,Chuang2016}. These biologically relevant species are referred to as interstellar complex organic molecules (COMs; see review paper: \citealt{Herbst2009,Herbst2017}). Additionally, methanol, a fully hydrogen-saturated derivative of CO, is also regarded as a starting point in the formation of COMs upon “energetic” processing via the interaction of UV-photons, cosmic particles, and X-rays \citep{Gerakines1996,Moore1996,Garrod2006,Oberg2009a,Modica2010,Chen2013,Boamah2014,Maity2015,Henderson2015,Paardekooper2016a,Chuang2017,Rothard2017}. 
These solid-state laboratory studies of energetic and nonenergetic scenarios all suggest an icy origin of COMs in star-forming regions. Systematic experimental studies also show that the corresponding relative yields of these complex species depend on the involved chemical triggers, the initial ice composition and substrate material underneath the ices \citep{Chuang2017,Ciaravella2018}. 

Furthermore, these newly formed species may participate in gas-phase reactions, once they or their fragments are released from the ice \citep{Balucani2015,Bertin2016,Taquet2016}. Several mechanisms have been studied in the recent past to understand the transfer from solid state material into the gas phase. Thermal desorption is well characterized for many different ices \citep{Brown2007,Burke2010}, but in dense dark molecular clouds, thermal desorption is highly unlikely due to the very low temperatures. Therefore, several nonthermal desorption mechanisms (e.g., photodesorption, chemical desorption, and ion sputtering) have been investigated to explain the net transfer from the solid state into the gas phase \citep{Garrod2007,Andersson2008,vanHemert2015,Bertin2016,Cruz-diaz2016,Minissale2016b,Chuang2018a,Oba2018,Dartois2019}. The resulting numbers, do not provide conclusive efficiencies so far.

COM gas phase signatures have been unambiguously detected in various astronomical environments from massive sources in the Galactic Center to comets in our solar system (\citealt{Herbst2009,Biver2014,Altwegg2017,Herbst2017} and references therein). 
Among $\sim$70 different COMs, which have been identified in the ISM, O-bearing COMs can conveniently be divided into two categories according to their chemical formulas; (1) C$_2$H$_\text{n}$O$_2$, such as methyl formate (HCOOCH$_3$), glycol-aldehyde (HCOCH$_2$OH), and ethylene glycol (HOCH$_2$CH$_2$OH), and (2) C$_2$H$_\text{n}$O, such as ketene (CH$_2$CO), acetaldehyde (CH$_3$CHO), ethanol (CH$_3$CH$_2$OH), and dimethyl ether (CH$_3$OCH$_3$). Such classification hints for possible chemical correlations among these species in terms of the degree of hydrogenation (i.e., n=2, 4, and 6). An example is the chemical connection between glycolaldehyde and ethylene glycol, which have commonly been found in solar-mass protostars, such as IRAS 16923-2422, NGC1333 IRAS 2A, and NGC1333 IRAS 4A \citep{Jorgensen2012,Jorgensen2016,Taquet2015,Coutens2015,Rivilla2017}. The consecutive hydrogenation of glycolaldehyde forming ethylene glycol has been experimentally confirmed on grain surfaces under molecular cloud conditions \citep{Fedoseev2017}. A similar chemical link can also be seen for species, such as ketene, acetaldehyde, and ethanol, which have been observed altogether toward several protostellar sources, such as NGC 7129 FIRS 2, SVS13-A, IRAS 16923-2422B, and L1157-B1 shock regions \citep{Bisschop2007,Fuente2014,Lefloch2017,Bianchi2018}. These species are topic of the present study.

A primary focus in this work is the origin of acetaldehyde in solid state COM chemistry. Acetaldehyde and its fully hydrogen-saturated species, ethanol, have been tentatively identified in interstellar ice mantles through their C-H deformation transitions at $\sim$7.41 and $\sim$7.25 $\mu$m, respectively \citep{Schutte1999,Oberg2011b}. Moreover, acetaldehyde is one of the most commonly detected COMs in massive hot cores and giant molecular clouds, such as Orion KL, Sgr B2N, and TMC-1, with a relatively high fractional abundance of $\chi$$_\text{H$_2$}$$\approx$10$^{-10}$-10$^{-9}$  \citep{Matthews1985,Turner1991,Ikeda2001,Charnley2004}. The isomers of acetaldehyde, (anti)syn-vinyl alcohol (CH$_2$CHOH), and ethylene oxide (\textit{c}-C$_2$H$_4$O) have higher enthalpies by (62.7)72.4 and 96 kJ mol$^{-1}$, respectively, than acetaldehyde. Both isomers have been identified in Sgr B2N by \cite{Turner2001}. Ethylene oxide has been reported in the PILS survey\footnote{http://youngstars.nbi.dk/PILS} of the low-mass protostar IRAS 16293-2422 by \cite{Lykke2017}, but a recent observational search for vinyl alcohol toward multiple solar-type protostars only resulted in upper limits \citep{Melosso2019}. Besides for high-mass protostars, the detection of acetaldehyde, usually together with ketene, has been reported toward solar-type dense molecular clouds, such as B1-b, L1689B, and B5 \citep{Bisschop2008,Oberg2010,Cernicharo2012,Bacmann2012,Taquet2017} and detections even have been reported for translucent clouds, such as CB 17, CB 24, and CB 228, with a surprisingly high $\chi$$_\text{H$_2$}$$\approx$10$^{-8}$  at low temperatures \citep{Turner1999}. These observations hint at a cold formation pathway of acetaldehyde and chemical derivatives (e.g., isomers and (de)hydrogenated products), during the very early stage of molecular clouds that is governed by H$_2$O ice formation, namely, before CO accretion starts. The goal of the present study is to investigate this pathway under controlled laboratory conditions. 

The formation of acetaldehyde and its isomers has been proposed to take place through either gas-phase or solid-state chemistry in star-forming regions. However, in the gas-phase routes, the model derived formation yields of C$_2$H$_4$O are usually (1-2 orders of magnitude) lower than the observed abundances \citep{Herbst1989,Lee1996}. Moreover, the proposed reaction scheme was questioned because the branching ratios of the associative products are unknown \citep{Charnley2004}. 
Therefore, multiple solid-state formation routes have been investigated not only to explain the observed abundances of C$_2$H$_4$O species, but also to incorporate acetaldehyde formation routes with its chemical relatives (e.g., ethanol and ketene), into a more comprehensive chemical network \citep{Charnley1997,Bennett2005a,Oberg2009a,Lamberts2019}. 

\begin{table*}[t]
	\caption{Summary of IR band strength and ionization cross-section values for species analyzed in this work.}             
	\label{tab.1}      
	\centering                          
	\begin{tabular}{llcllcrr}
		\hline
		\hline
		Species & Formula & IR position & \textit{A} in transmission & \textit{A} used in this work\tablefootmark{c} & Ref. & Ionization $\sigma$ & Ref. \\
		& & (cm$^{-1}$) & (cm molecule$^{-1}$) & (cm molecule$^{-1}$) & & ($\AA^2$) $\sigma$$_{70eV}$/$\sigma$$_{max(cal.)}$ & exp./cal. \\
		\hline
		Ketene & CH$_2$CO & 2133 & 1.2$\times$10$^{-16}$ & 6.6$\times$10$^{-16}$ & [1] & - /6.5 & - /[5] \\
		Acetaldehyde & CH$_3$CHO & 1705 & 3.0$\times$10$^{-17}$\tablefootmark{a} & 1.7$\times$10$^{-16}$ & [2] & 6.5/6.7 & [6]/[5] \\
		Vinyl alcohol & CH$_2$CHOH & 1639 & 2.8$\times$10$^{-17}$\tablefootmark{a} & 1.5$\times$10$^{-16}$ & [2] & - /6.5 & - /[7] \\
		Ethanol & CH$_3$CH$_2$OH & 1041 & 1.4$\times$10$^{-17}$ & 7.7$\times$10$^{-17}$ & [3] & 7.3/7.5 & [8]/[9] \\
		Acetic acid & CH$_3$COOH & ~1742 & 1.1$\times$10$^{-16}$\tablefootmark{b} & 6.1$\times$10$^{-16}$ & [4] & 7.7/7.5 & [10]/[5] \\
		\hline
	\end{tabular}
	\tablefoot{\tablefootmark{a}{Calculated value.} 
		\tablefootmark{b}{Gas-phase value.}
		\tablefootmark{c}{Obtained from literature value (transmission) by multiplying a conversion factor of 5.5 for the current RAIRS geometry setting.}
	}
	\tablebib{
		[1] \cite{Berg1991};
		[2] \cite{Bennett2005a}; 
		[3] \cite{Hudson2017}; 
		[4] \cite{Marechal1987};
		[5] \cite{Gussoni1998}; 
		[6] \cite{Bull2008}; 
		[7] NIST database;
		[8] \cite{Hudson2003}; 
		[9] \cite{Gray1984}; 
		[10] \cite{Bhutadia2012}.}
\end{table*}

The $\sim$3.4 $\mu$m IR features that have commonly been observed toward diffuse clouds are widely accepted originating from hydrocarbons attached to amorphous carbonaceous dust grains (HAC, \citealt{Sandford1991}). As a consequence of exposing bare grains to energetic particles and photons, these aliphatic functional groups are expected to be destroyed leading to small molecules or even unsaturated fragments \citep{Teyssier2004,Alata2014}. This top-down processing is generally considered to enrich the hydrocarbon abundance in the ISM. For example, a solid-state origin of C$_2$H$_2$ has been proposed to explain the abundant C$_2$H$_2$ in the gas phase toward interstellar clouds (C$_2$H$_2$/CO$\cong$10$^{-3}$) and massive YSOs (C$_2$H$_2$/H$_2$O$\cong$(1-5)$\times$10$^{-2}$) \citep{Lacy1989,Lahuis2000}. Modeling and laboratory studies have also shown an efficient erosion of hydrogenated amorphous carbonaceous dust (HAC) triggered by energetic sources and photodissociation of ionized polycyclic aromatic hydrocarbons (PAHs) forming C$_2$H$_2$, C$_2$H, and C$_2$ \citep{Jochims1994,Allain1996,LePage2003,Jager2011,Zhen2014,West2018}. The resulting loss channel strongly depends on the size of the parent compound and the applied energy. In molecular clouds as the dust temperature decreases, C$_2$H$_2$ (and its radicals) may further associate with impacting H-atoms to form (semi-)saturated C$_2$H$_4$ and C$_2$H$_6$ on dust grains \citep{Tielens1992,Hiraoka2000,Kobayashi2017}. These hydrocarbons could act as an alternative starting point for the formation of organic species by interacting with other accreting atoms, such as O-atoms. Based on this, \cite{Tielens1992} proposed a series of surface reactions including several atom addition reactions starting from C$_2$H$_2$ ice (see also figure 2 in \citealt{Charnley2004});
\begin{subequations}
\begin{equation}\label{Eq2.1}
\text{C$_2$H$_2$}\xrightarrow{\text{+H}}\text{C$_2$H$_3$}\xrightarrow{\text{+O}}\text{C$_2$H$_3$O}\xrightarrow{\text{+H}}\text{C$_2$H$_4$O}$ $ \text{isomers};
\end{equation}
\begin{equation}\label{Eq2.2}
\text{CH$_3$CHO} \text{ or } \text{CH$_2$CHOH}\xrightarrow{\text{+2H}}\text{CH$_3$CH$_2$OH}.
\end{equation}
\end{subequations}
Experimental studies have shown that the surface chemistry between C$_2$H$_4$ and suprathermal (“hot”) O-atoms ($^1$D and $^3$P) generated by electron or UV-photon impact can explain the production of C$_2$H$_4$O isomers and chemically related species in the ISM \citep{DeMore1969,Hawkins1983,Bennett2005b,Bergner2019}. However, it is very difficult to clarify the dominant formation pathway of COMs in such solid-state reactions triggered by energetic processes. For example, besides O-atom addition reactions of hydrocarbons, H-atom abstraction reactions induced by O-atoms produced by photodissociation (i.e, suprathermal) are expected to result in the production of reactive hydroxyl radicals (OH). These newly formed OH-radicals could then react with hydrocarbons or their photofragments (i.e., C$_2$H$_\text{2n-1}$), additionally contributing to COM formation as suggested in previous experimental and theoretical work \citep{Moore1998,Hudson2003,Basiuk2004}. Moreover, \cite{Ward2011} used a microwave-discharge atom beam with a quenching PTFE tube to study the “thermal” O-atom ($^3$P) reaction with C$_2$H$_4$ mainly leading to the formation of ethylene oxide, instead of efficiently resulting in acetaldehyde, which is the dominant C$_2$H$_4$O species in the ISM.

The present study is motivated by the abundant detection of acetaldehyde in (translucent and dense) molecular clouds, where nonenergetic processes predominate surface chemistry and the majority of O-atoms is efficiently converted to OH-radicals and H$_2$O \citep{Cuppen2007,vanDishoeck2013}. It is important to note that in the early stage of molecular clouds, cosmic rays and UV photons are still present and trigger energetic chemistry in the gas phase, which could help in producing more atoms and radicals available to participate surface reactions. As reported in previous ab initio theoretical calculations and experimental studies, a small barrier or barrierless reaction of adding OH-radicals to C$_2$H$_2$ offers a new solid-state pathway to form interstellar COMs by incorporating unsaturated hydrocarbons as molecular backbone \citep{Miller1989,Basiuk2004,McKee2007}. In this work, the case study of molecule-radical (unsaturated hydrocarbon-OH radical) reactions leading to COMs is expected to be a general and very efficient mechanism, since these routes only require the formation of OH-radicals, which are the intermediate species leading to H$_2$O, next to the unsaturated hydrocarbon. A very recent experimental study by \cite{Qasim2019a} has demonstrated propanol formation through a similar mechanism involving propyne (C$_3$H$_4$) and OH-radicals. In this study, we aim at extending the proposed solid-state network and investigating the formation of commonly detected acetaldehyde and related species through the surface reactions of C$_2$H$_2$ with OH-radicals and H-atoms in the early stage of molecular clouds, where H$_2$O ice starts forming on dust grains. 

\section{EXPERIMENTAL}
\begin{table*}[t!]
	\caption{Overview of the performed experiments.}             
	\label{tab.2}      
	\centering                          
	\begin{tabular}{clcccccc}
		\hline
		\hline
		No. & Experiment & T & Flux(C$_2$H$_2$) & Flux(O$_2$) & Flux(H) & Ratio & Time \\
		& & (K) & (molecules cm$^{-2}$s$^{-1}$) & (molecules cm$^{-2}$s$^{-1}$) & (atoms cm$^{-2}$s$^{-1}$) & & (min.) \\
		\hline
		exp. 1 & C$_2$H$_2$ & 10 & 3$\times$10$^{11}$ & - & - & - & 360 \\
		exp. 2 & C$_2$H$_2$+H & 10 & 3$\times$10$^{11}$ & - & 1.2$\times$10$^{13}$ & 1: - :40 & 360 \\
		exp. 3 & C$_2$H$_2$+O$_2$+H & 10 & 3$\times$10$^{11}$ & 3$\times$10$^{12}$ & 1.2$\times$10$^{13}$ & 1:10:40 & 60\&360 \\
		exp. 4 & C$_2$H$_2$+$^{18}$O$_2$+H & 10 & 3$\times$10$^{11}$ & 3$\times$10$^{12}$ & 1.2$\times$10$^{13}$ & 1:10:40 & 360 \\
		\hline
		No. & Control experiment & T & Flux(CH$_3$CHO) & Flux(O$_2$) & Flux(H) & Ratio & Time \\
		& & (K) & (molecules cm$^{-2}$s$^{-1}$) & (molecules cm$^{-2}$s$^{-1}$) & (atoms cm$^{-2}$s$^{-1}$) & & (min.) \\
		\hline
		exp. c1 & CH$_3$CHO(pre.)+H & 10 & - & - & 1.2$\times$10$^{13}$ & - & 30 \\
		exp. c2 & CH$_3$CHO+H & 10 & 6$\times$10$^{11}$ & - & 1.2$\times$10$^{13}$ & 2: - :40 & 60\&360 \\
		exp. c3 & CH$_3$CHO+O$_2$+H & 10 & 6$\times$10$^{11}$ & 3$\times$10$^{12}$ & 1.2$\times$10$^{13}$ & 2:10:40 & 60\&360 \\
		\hline
	\end{tabular}
\end{table*}

All experiments were performed by using an ultra-high vacuum (UHV) setup, SURFRESIDE$^2$, which has been designed and fully optimized for studies of atom- and radical-addition reactions between thermalized species (i.e., for nonenergetic processing). Details of operation and calibration have been described in \cite{Ioppolo2013}, and the latest update is available in \cite{Fedoseev2016} and \cite{Chuang2018a}. Here only the relevant details are listed. The studied ices were grown on a gold-plated copper substrate mounted on the tip of a closed-cycle helium cryostat that is positioned in the center of the UHV chamber. Base pressures were typically around 5$\times$10$^{-10}$ mbar. The substrate temperature was monitored by two silicon diode sensors with 0.5 K accuracy and can be regulated in the range between 8 and 450 K by means of a resistive cartridge heater. Gaseous species, namely O$_2$ (Linde 5.0 and $^{18}$O$_2$ from Campro Scientific 97\%), C$_2$H$_2$ (Praxair; 5\% in Helium), and vapor of CH$_3$CHO (Aldrich $\geq$99.5\%), which is purified through multiple freeze-pump-thaw cycles, were admitted into the UHV chamber through two high-precision leak valves and deposited onto the precooled substrate at 10 K. 

The ices were grown with submonolayer precision (where 1 ML is assumed to be 1$\times$10$^{15}$ molecules cm$^{-2}$) with typical deposition rates of $\sim$3$\times$10$^{12}$, $\sim$3$\times$10$^{11}$, and $\sim$6$\times$10$^{11}$ molecules cm$^{-2}$s$^{-1}$ for O$_2$, C$_2$H$_2$, and CH$_3$CHO, respectively. Because O$_2$ is IR inactive, the flux obtained by the Langmuir estimation must be treated as an upper limit. In this work, H-atoms with a flux of $\sim$1.2$\times$10$^{13}$ atoms cm$^{-2}$s$^{-1}$ were introduced either from a Hydrogen Atom Beam Source (HABS; \citealt{Tschersich2000}) or Microwave Atom Source (MWAS; \citealt{Anton2000}). Each atom beam line was differentially pumped with a base pressure of $\leq$1$\times$10$^{-9}$ mbar and connected to the UHV chamber through a shutter valve and a bended quartz pipe, which was used to efficiently quench and thermalize excited H-atoms and nondissociated molecules through multiple collisions with the pipe wall, before these impact onto the ice substrate. Application of these two distinct H-atom sources did not result in clear experimental differences.

In this study, the applied fluxes and deposition ratios are chosen such that these allow one to investigate the proposed COM formation mechanisms, rather than trying to fully simulate astrochemical conditions in molecular clouds (e.g., the species accretion rate). The ratio of C$_2$H$_2$ to H-atom used in this study followed the trend of astronomical observations of C$_2$H$_2$; the gaseous abundance of C$_2$H$_2$ is a few orders of magnitude lower than the corresponding H-atom value, assuming that the abundance ratio of CO to H-atom is unity \citep{Lacy1989,Evans1991}. O$_2$ is intentionally used as a precursor to efficiently produce OH-radicals through consequent reactions, that is, \text{O$_2$}$\xrightarrow{\text{H}}$\text{HO$_2$}$\xrightarrow{\text{H}}$\text{$2$OH}, with negligible (or even no) barrier \citep{Cuppen2010,Lamberts2013}. Traces of O-atoms can be formed under our experimental conditions through three minor reaction channels, namely \text{H + HO$_2$}$\longrightarrow$\text{H$_2$O + O}, \text{OH + OH}$\longrightarrow$\text{H$_2$O + O}, and \text{OH + O$_2$}$\longrightarrow$\text{HO$_2$ + O} \citep{Cuppen2010}. The formed O-atoms are then expected to react with abundant and highly mobile H-atoms resulting in more OH-radicals available in the ice mixture via barrierless reactions. Products, such as H$_2$O and H$_2$O$_2$, are also expected to form on the grain surface along with COMs, representing the H$_2$O-rich environment in translucent clouds. The comprehensive formation network leading to H$_2$O ice has been extensively investigated via the solid-state interactions between H-atoms and oxygen allotropes (i.e., O, O$_2$, and O$_3$). The interested reader is referred to review papers by \cite{vanDishoeck2013} and \cite{Linnartz2015}.
 
Both the predeposition method, that is, H-atom accretion on top of preexisting ices, and the codeposition method, that is, a simultaneous deposition of ices with H-atoms, were used in this work. The former one is typically utilized to obtain the kinetics of reactions, and the latter one can overcome the limitation of restricted H-atom penetration depth into the bulk ice (i.e., few upper ice layers) \citep{Watanabe2008,Chuang2018a}. Moreover in molecular clouds, atoms and molecules are expected to continuously accrete onto grains from the gas phase, which is simulated by the codeposition experiments \citep{Fedoseev2015b}. The ice sample was monitored in situ by Fourier transform reflection absorption infrared spectroscopy (RAIRS) in the range from 700 to 4000 cm$^{-1}$, with 1 cm$^{-1}$ resolution. 

The column density and composition of the ices were obtained by the modified Beer–Lambert law converting IR absorbance area to absolute abundance. Absorption band strength values (i.e., the so-called \textit{A} value) of parent and product species have been reported in the literature for ice transmission measurements for quite some species. However, the band strength values of vinyl alcohol and acetaldehyde were taken from the theoretical values reported in \cite{Bennett2005b}. Here, we assumed that the band strength ratio between “transmission mode” and “reflection mode” is constant for all studied species \citep{Oberg2009a,Chuang2018b}. The averaged conversion factor, namely \textit{A}$_{\text{reflection}}$$\diagup$\textit{A}$_{\text{transmission}}$= 5.5$\pm$0.8, for the present IR geometry setting, was obtained by comparing the \textit{A} values reported in previous RAIR measurements by \cite{Chuang2018a} to the values (in transmission) published by \cite{Bouilloud2015} for three well-studied interstellar ice analogs, that is, CO, H$_2$CO, and CH$_3$OH. Both measurements used the same method, namely laser interference (at 632.8 nm), and the same physical parameters for ice structure, such as ice density and refractive index, to determine the absolute ice thickness. As a consequence, the relative uncertainties of the conversion ratio obtained from these two measurements could be largely minimized.

\begin{figure*}[t!]
	\begin{center}
		\includegraphics[width=\textwidth]{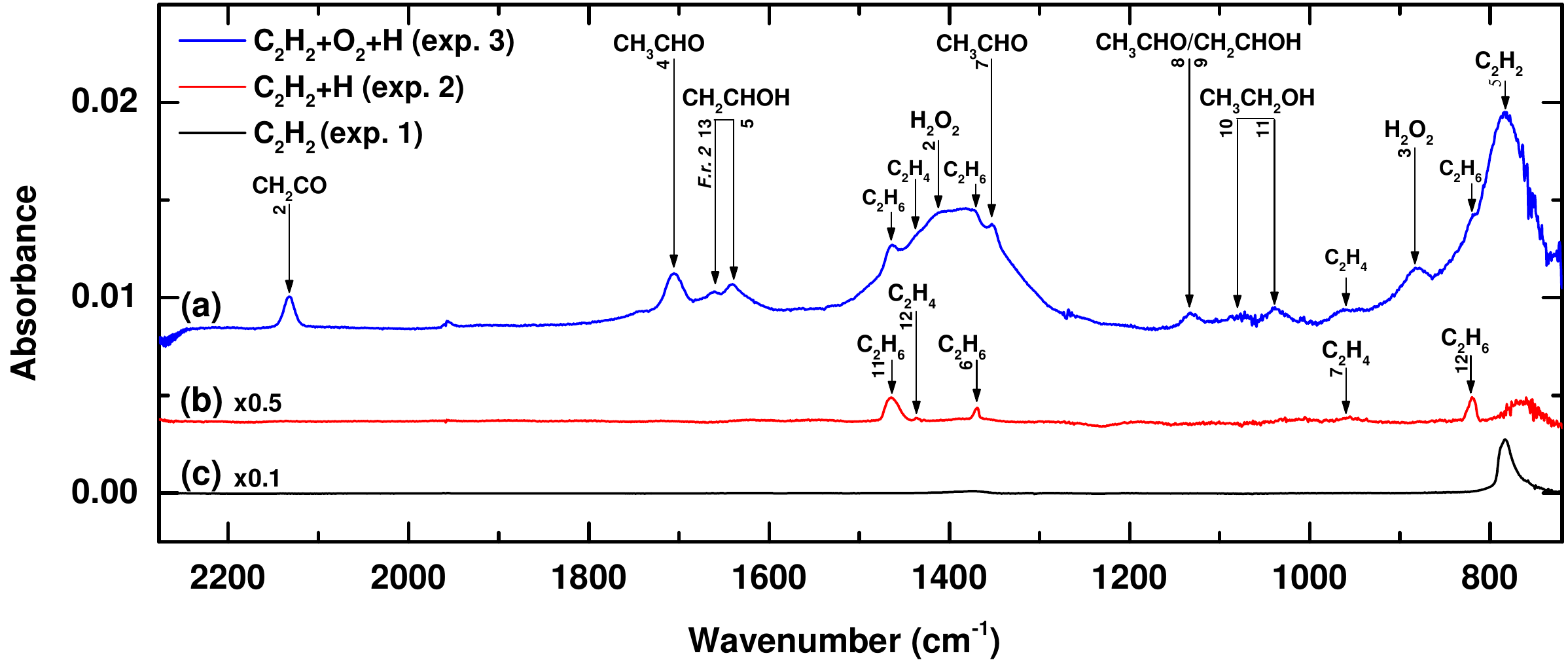}
		\caption{IR spectra obtained after 360 min of codeposition of (a) C$_2$H$_2$ + O$_2$ + H (exp. 3), (b) C$_2$H$_2$ + H (exp. 2), and (c) reference experiment of C$_2$H$_2$ (exp. 1) at 10 K. The applied deposition ratio of C$_2$H$_2$:(O$_2$ ):H was 1:(10):40, and the used H-atom flux was 1.2$\times$10$^{13}$ atoms cm$^{-2}$s$^{-1}$. IR spectra are offset for clarity}
		\label{Fig1}
	\end{center}
\end{figure*}

After completion of the deposition experiment, a temperature-programmed desorption experiment using a Quadrupole mass spectrometer (TPD-QMS) was performed with a ramping rate of 5 K min$^{-1}$. The initial and newly formed species thermally desorb at a certain substrate temperature, depending on the species’ binding energy with surrounding molecules. These species largely fragment, exhibiting a unique ionization fragmentation pattern upon electron impact (at 70 eV). The combined information of desorption temperature and fragmentation pattern is a vital diagnostic tool to securely identify the species produced in the solid-state by comparing with standard samples in the NIST database. These characteristics can also be applied to distinguish isomers, which share the same molecular mass channel in the QMS spectrum but typically have different desorption temperatures and fragmentation patterns. Furthermore, the integral of a species desorption profile in TPD-QMS spectra followed by a series of calibrations including fragmentation fraction (\textit{F}$_{\text{f}}$(m)), sensitivity fraction of the QMS (\textit{S}(m/z)) and total ionization cross-section value (\textit{$\sigma$}$^+$(mol) in Å$^2$) can be used to determine a species relative abundance \citep{Martin-Domenech2015,Chuang2018thesis}. All relevant fragmentation values can be found in the NIST database, except for vinyl alcohol (CH$_2$CHOH), for which values are available from \cite{Turecek1984}. Ionization cross-section values have been reported in experimental studies. For those that have no experimental value, such as ketene and vinyl alcohol, an empirical correlation method, that is, $\sigma$$_\text{max}$$\cong$c$\cdot$$\alpha$ ($\sigma$$_\text{max}$: maximum ionization cross-section, c: correlation constant of 1.48 in Å$^{-1}$ and $\alpha$: molecular polarizability volume in Å$^3$), was used \citep{Lampe1957,Bull2011}. For organic molecules, such as aldehydes and alcohols, the maximum ionization cross-section value is typically located at 90$\pm$10 eV and the intensity differences between 70 and 90 eV are very small ($<$5\%) \citep{Hudson2003,Bull2008}. 

In addition to the QMS measurements, desorbing species during TPD experiments can also be monitored by acquiring the peak depletion in RAIR spectra (TPD-IR) obtained before and after ice constituent sublimation. The RAIR band strength and QMS ionization cross-section values of relevant species studied in this work are summarized in Table \ref{tab.1}. The derived molecular (relative) abundances can easily be recalibrated in the future as more precise values become available. The error bars of IR and QMS data are derived from the feature’s Gaussian fitting and from the integrated noise signal of a blank experiment in the same temperature range for each mass signal (as instrumental errors), respectively. They do not account for uncertainties resulting from the baseline subtraction procedure and the uncertainties of band strength and ionization cross-section. Table \ref{tab.2} lists the relevant experiments performed in this work.

\section{RESULTS}
\subsection{COM formation in experiments: C$_2$H$_2$ + O$_2$ + H}
Figure \ref{Fig1} presents the IR absorption spectra obtained after codeposition of (a) C$_2$H$_2$ + O$_2$ + H (exp. 3), (b) C$_2$H$_2$ + H (exp. 2), both with an H-atom flux of 1.2$\times$10$^{13}$ atoms cm$^{-2}$s$^{-1}$, and (c) a reference experiment of C$_2$H$_2$ (exp. 1) at 10 K for 360 min. The applied deposition ratio of C$_2$H$_2$(:O$_2$):H was 1(:10):40. In exps. 2 and 3, the newly formed species ethylene (C$_2$H$_4$) and ethane (C$_2$H$_6$) are visible in the IR spectrum. IR features at 1437 and 956 cm$^{-1}$ originate from the CH$_2$ scissoring ($\nu$$_{12}$) and wagging ($\nu$$_7$) modes of C$_2$H$_4$, respectively \citep{Shimanouchi1972,Bennett2006}. The fully hydrogenated species, C$_2$H$_6$, can also be identified by spectral features at 
1464 (CH$_3$ asym. deformation ($\nu$$_{11}$)), 1371 (CH$_3$ sym. deformation ($\nu$$_6$)), and 822 (CH$_3$ rocking ($\nu$$_{12}$)) cm$^{-1}$ \citep{Bennett2006,Kaiser2014}. These experimental results are fully consistent with the previous investigations of C$_2$H$_2$ hydrogenation on grain surfaces at 10 K, following the reaction scheme of C$_2$H$_2$$\xrightarrow{\text{+2H}}$C$_2$H$_4$$\xrightarrow{\text{+2H}}$C$_2$H$_6$, in theoretical and laboratory studies \citep{Tielens1992,Hiraoka1999,Hiraoka2000,Kobayashi2017}. As discussed, intermediate radicals along the C$_2$H$_2$ hydrogenation scheme, such as C$_2$H$_3$ and C$_2$H$_5$, are not detected, indicating their low abundance due to the high reactivity of radicals compared to stable molecules. The derived abundance ratio of C$_2$H$_6$ to C$_2$H$_4$ in exp. 2, namely ~9, is very similar to the final ratio obtained by \citealt{Kobayashi2017} (at steady state in their figure 7) supporting the conclusion that the activation barrier of H-atom addition reaction to C$_2$H$_4$ is lower than that of C$_2$H$_2$ hydrogenation \citep{Charnley2004,Kobayashi2017}.

\begin{figure*}[t]
	\begin{center}
		\includegraphics[width=\textwidth]{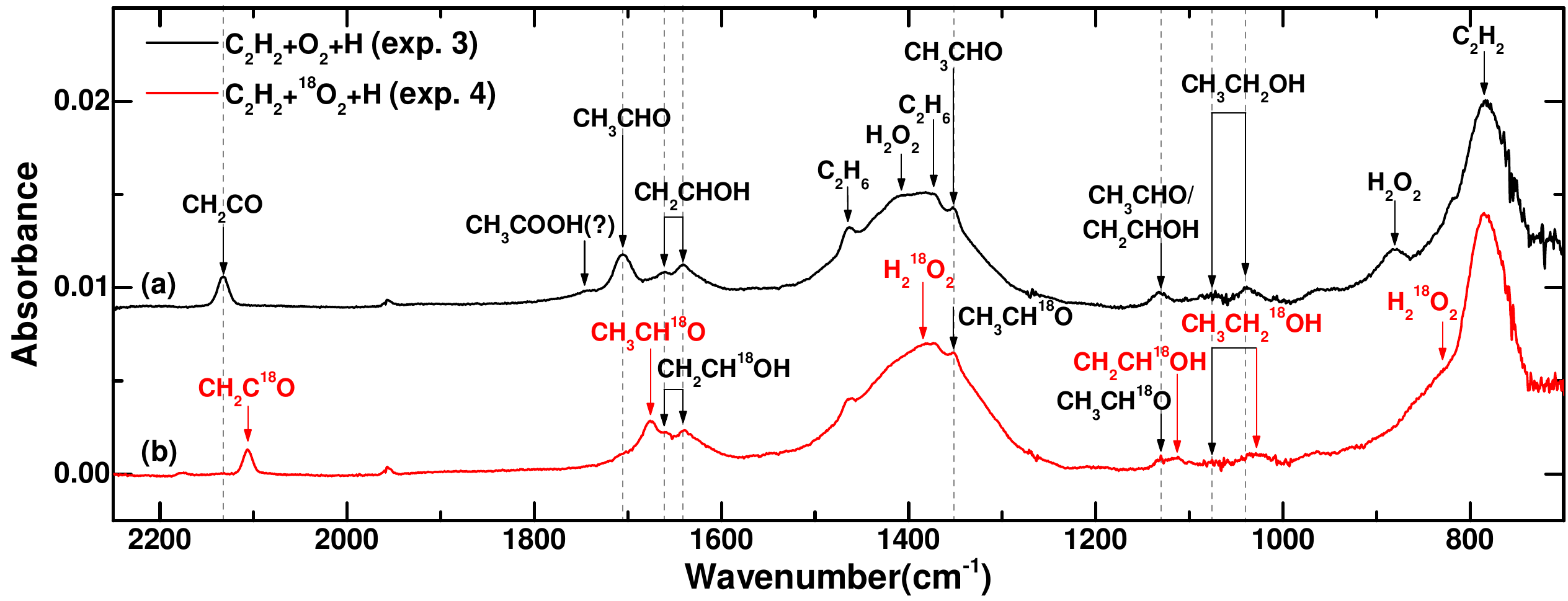}
		\caption{IR spectra obtained after the codepositions of (a) C$_2$H$_2$ + O$_2$ + H (exp. 3), and (b) C$_2$H$_2$ + $^{18}$O$_2$ + H (exp. 4) for 360 min at 10 K. The applied deposition ratio of C$_2$H$_2$:$^{(18)}$O$_2$:H was 1:10:40, and the used H-atom flux was 1.2$\times$10$^{13}$ atoms cm$^{-2}$s$^{-1}$ The dashed lines are shown for the peak positions of regular molecules. IR spectra are offset for clarity}
		\label{Fig2}
	\end{center}
\end{figure*}

In addition to the formation of these two hydrocarbons, in exp. 3, the surface chemistry of C$_2$H$_2$ involving OH-radicals, which are predominantly generated from O$_2$$\xrightarrow{\text{+H}}$HO$_2$$\xrightarrow{\text{+H}}$\text{2OH}, leads to the appearance of many other IR features originating from O-bearing species. The peaks at 1705, 1351, and 1132 cm$^{-1}$ are due to the C=O stretching ($\nu$$_4$), CH$_3$ (sym.) deformation ($\nu$$_7$), and C-C rocking ($\nu$$_8$) vibrational modes of acetaldehyde (CH$_3$CHO) \citep{Hollenstein1971,Hawkins1983,Bennett2005b,TerwisschavanScheltinga2018,Hudson2020}. The peak around 1132 cm$^{-1}$ could also have an additional contribution from the C-O stretching ($\nu$$_9$) mode of vinyl alcohol (CH$_2$CHOH). The other IR features of vinyl alcohol, overlapped with a broad H$_2$O feature, can be observed at 1639 cm$^{-1}$ due to C-C stretching ($\nu$$_5$) modes and at 1661 cm$^{-1}$ due to its Fermi resonance with the overtone of CH$_2$ wagging (2$\nu$$_{13}$) mode \citep{Rodler1984,Hawkins1983,Koga1991}. The IR features with relatively weak absorption coefficiency, which originate from the CH$_3$ deformation ($\nu$$_5$) mode of acetaldehyde and CH$_2$ scissor ($\nu$$_6$) mode of vinyl alcohol, are strongly blended with a broad H$_2$O$_2$ peak and hydrocarbon peaks around 1413 cm$^{-1}$. A detailed spectral deconvolution for IR features of acetaldehyde and vinyl alcohol around 1700 cm$^{-1}$ can be seen in Fig. \ref{Fig.A1}(a) in the Appendix. Ethanol (CH$_3$CH$_2$OH) is clearly observed at 1041 cm$^{-1}$ through its C-O stretching ($\nu$$_{11}$) mode along with a relatively small feature at 1086 cm$^{-1}$ ($\nu$$_{10}$; CH$_3$ rocking mode) \citep{Barnes1970,Mikawa1971,Boudin1998}. The peak located at 2133 cm$^{-1}$ can be assigned to the C=O stretching ($\nu$$_2$) mode of ketene (CH$_2$CO) \citep{Hudson2013}. A tiny peak at $\sim$1744 cm$^{-1}$ is tentatively assigned to the C=O stretching ($\nu$$_4$) of acetic acid \citep{Bahr2006,Bertin2011}.

An isotopic experiment of C$_2$H$_2$ + $^{18}$O$_2$  + H was performed to further secure the IR assignments of the O-bearing products. A direct comparison of IR spectra between experiments (a) C$_2$H$_2$ + O$_2$  + H (exp. 3) and (b) C$_2$H$_2$ + $^{18}$O$_2$ + H (exp. 4) is presented in Fig. \ref{Fig2} and clearly shows the spectral redshift corresponding to species’ C$^{18}$O vibrational modes. Besides H$_2$$^{18}$O$_2$ signals (i.e., 1384 and 832 cm$^{-1}$), IR features of CH$_2$C$^{18}$O ($\nu$$_2$), CH$_3$CH$^{18}$O ($\nu$$_4$), CH$_2$CH$^{18}$OH ($\nu$$_9$), and CH$_3$CH$_2$$^{18}$OH ($\nu$$_{11}$) are found at 2106, 1677, 1114, and 1030 cm$^{-1}$, respectively \citep{Hawkins1983,Rodler1984,Hudson2013,Maity2014,Bergner2019}. Peak positions due to transitions not directly affected by O-atoms, such as the C-C stretching modes (1662 and 1639 cm$^{-1}$) of CH$_2$CH$^{18}$OH and the CH$_3$ deformation (1351 cm$^{-1}$) as well as the C-C stretching (1133 cm$^{-1}$) mode of CH$_3$CH$^{18}$O do not shift more than the 1 cm$^{-1}$ with respect to the same transitions of the main isotope, which is within our spectral resolution \citep{Rodler1984,Hawkins1983}. The (non)peak shifts in the isotopic control experiment greatly support the IR feature assignments of acetaldehyde, vinyl alcohol, ketene, and ethanol, which are the newly formed O-bearing species from the interactions between C$_2$H$_2$ and OH-radicals as well as H-atoms at 10 K.

\begin{figure}[b!]
	\vspace{-0.3cm}
	\begin{center}
		\includegraphics[width=90mm]{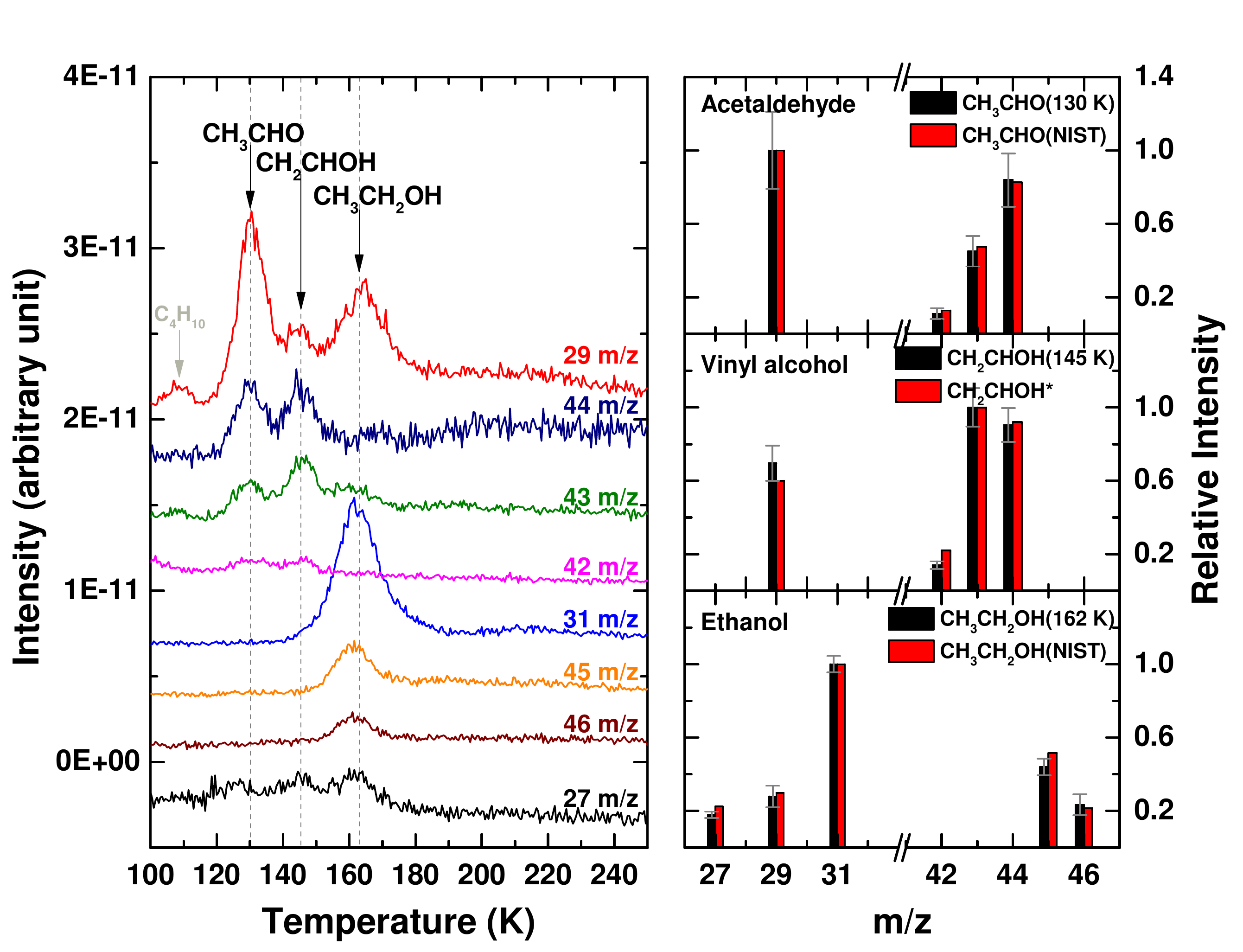}
		\caption{Left: TPD mass spectra obtained after 60 min of codeposition of C$_2$H$_2$ + O$_2$  + H (exp. 3) at 10 K. The applied deposition ratio of C$_2$H$_2$:O$_2$:H was 1:10:40, and the used H-atom flux was 1.2$\times$10$^{13}$ atoms cm$^{-2}$s$^{-1}$. Only relevant m/z channels are shown and shifted for clarity. The arrows and dashed lines indicate the peak position of the corresponding COMs. Right: comparison of the ionization fragmentation pattern of the detected COMs (black) with available values in the literature (red; NIST database: acetaldehyde and ethanol; \cite{Turecek1984}: vinyl alcohol). }
		\label{Fig3}
	\end{center}
\end{figure}

\begin{figure}[t]
	\begin{center}
		\includegraphics[width=90mm]{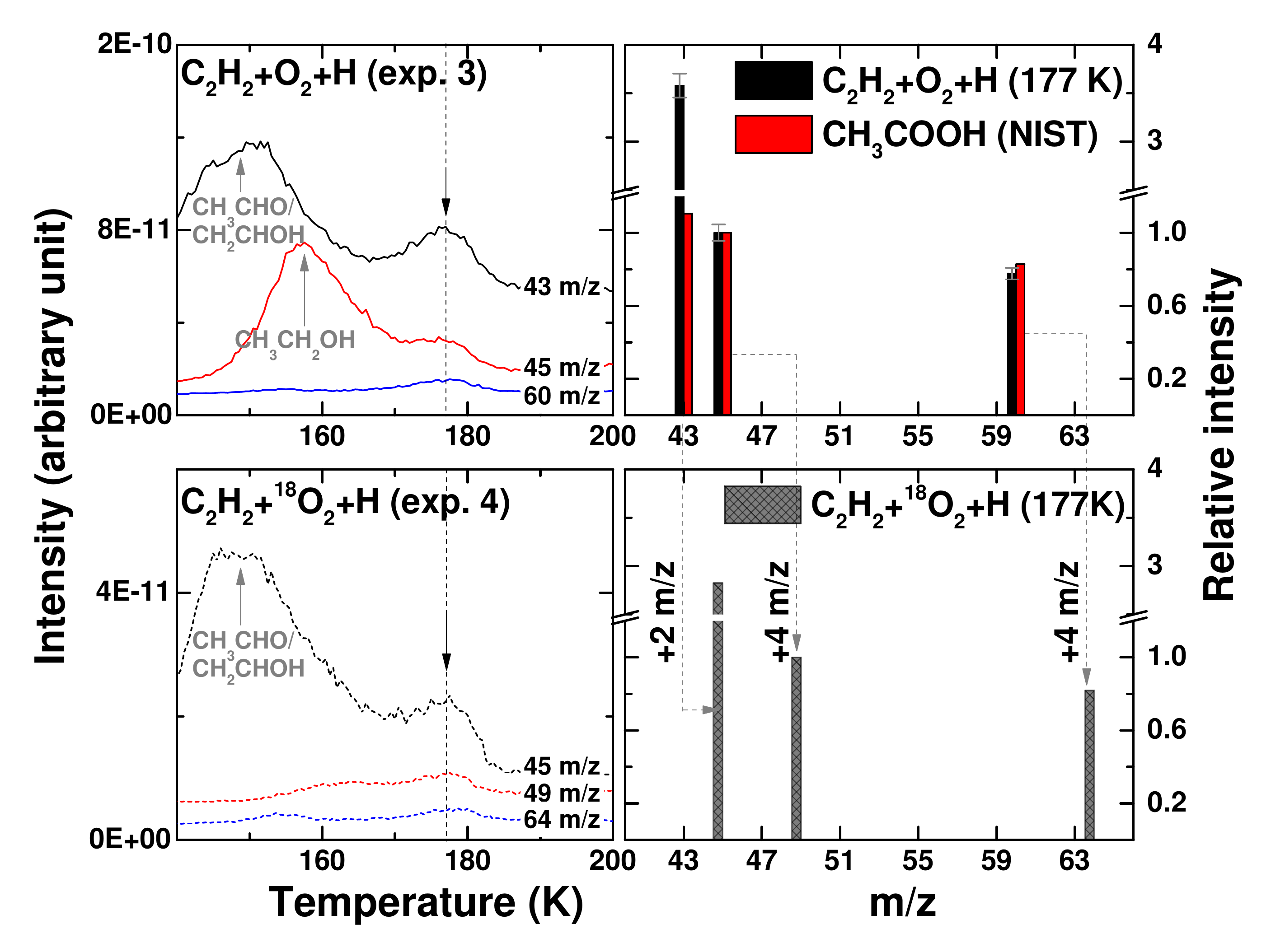}
		\caption{Left: TPD mass spectra obtained after 360 min of codeposition of C$_2$H$_2$ + O$_2$ + H (exp. 3) and C$_2$H$_2$ + $^{18}$O$_2$ + H (exp. 4) at 10 K. The applied deposition ratio of C$_2$H$_2$:$^{(18)}$O$_2$ :H was 1:10:40, and the used H-atom flux was 1.2$\times$10$^{13}$ atoms cm$^{-2}$s$^{-1}$. Only relevant m/z channels are shown and shifted for clarity. The dashed line indicates the peak position of the corresponding COMs. Right: comparison of the ionization fragmentation pattern for the QMS peak at 177 K (black) with data of CH$_3$COOH in NIST (red). The position of the dashed arrows present for the isotope mass signal shifts in exp. 4.}
		\label{Fig4}
	\end{center}
\end{figure}

Figure \ref{Fig3} presents TPD-QMS spectra obtained after codeposition of C$_2$H$_2$ + O$_2$ + H (exp. 3) at 10 K for relevant m/z signals, which are selected to further pinpoint the formation of acetaldehyde, vinyl alcohol, and ethanol. Here, in order to avoid severe overlap of product peaks in a small temperature range (i.e., 120-180 K), the QMS data are derived from a 60-min only codeposition experiment. The typical QMS spectra for 360-min experiments are presented in Fig. \ref{Fig4} for isolated desorption peaks at higher sublimation temperatures. As described in the Experimental Section, TPD-QMS provides two additional tools to identify newly formed products, that is, via the desorption temperature and via the characteristic dissociative ionization fragmentation pattern. In the left-hand panel of Fig. \ref{Fig3}, three main desorption peaks are located at 130, 145, and 162 K. The first two peaks at 130 and 145 K consist of the same mass fragments (e.g., 29, 42, 43, and molecular mass 44 m/z), but with different relative intensities (see right panel), reflecting that they originate likely from different chemical isomers. Based on the available desorption temperature data of C$_2$H$_4$O isomers reported by \cite{Abplanalp2016}, we conclude that the desorbing peak at 130 K originates from acetaldehyde and the peak at 145 K is due to vinyl alcohol. A fully consistent desorption behavior visualized by FTIR signals of acetaldehyde and vinyl alcohol can be found during a TPD experiment in Fig. \ref{Fig.A1}(b) in the Appendix. The third peak in Fig. \ref{Fig3} (i.e., mass signals of 27, 29, 31, 45, and 46 m/z) around 162 K can be assigned to ethanol \citep{Oberg2009a,Bergantini2017}. The mass fragmentation patterns of three products obtained by integrating the QMS desorption profiles are presented in the right-hand panel of Figure \ref{Fig3}. These patterns are generally consistent with the standard electron (70 eV) ionization pattern as available from the NIST database (for acetaldehyde and ethanol) and \cite{Turecek1984} (for vinyl alcohol). The experimental results of the QMS spectra during the TPD experiment are in a good agreement with the above IR spectral findings. 

An experiment with a longer codeposition time is expected to result in a higher abundance of products while still keeping the same ice composition ratio. The linearly proportional correlation between experimental time and formation yield can be used to identify higher-generation products, typically with much smaller abundance compared to the first-generation species, by extending the codeposition time. Figure \ref{Fig4} shows the QMS spectra obtained after codeposition of C$_2$H$_2$ + O$_2$ + H (exp. 3) and C$_2$H$_2$ + $^{18}$O$_2$ + H (exp. 4) for 360 min. Besides the desorption peaks previously assigned to acetaldehyde and vinyl alcohol, which are now difficult to be deconvoluted, as well as ethanol, a new desorption peak can be found around 177 K showing mass signals of 43, 45, and 60 m/z in experiment C$_2$H$_2$ + O$_2$ + H (exp. 3). The observed desorption temperature and the total molecular mass (60 m/z) suggest that this signal is due to the direct sublimation of acetic acid (CH$_3$COOH) or its thermal codesorption with H$_2$O$_2$ or H$_2$O \citep{Bahr2006,Oberg2009a,Bertin2011}. In the isotope experiment C$_2$H$_2$ + $^{18}$O$_2$ + H (exp. 4), the corresponding mass signal shifts; the signals at 43 m/z (CH$_3$CO$^+$), 45 m/z (COOH$^+$), and 60 m/z (CH$_3$COOH$^+$) are accordingly shifted to mass fragments of 45 m/z (CH$_3$C$^{18}$O$^+$), 49 m/z (C$^{18}$O$^{18}$OH$^+$), and 64 m/z (CH$_3$C$^{18}$O$^{18}$OH$^+$), respectively. It should be noted that the relative intensities of 43 and 45 m/z in exps. 3 and 4 are too high for the standard fragmentation pattern of acetic acid (NIST database). This could be due to a high baseline (i.e., high concentration) of background mass signal at 43 m/z (or 45 m/z in the isotope experiment) originating from residual acetaldehyde and vinyl alcohol in the chamber. However, the ratio of mass signals between 45 and 60 m/z (i.e., 0.78), which have a lower signal baseline, is comparable with the standard ratio from the NIST database (i.e., 0.82). A comparable value in the corresponding isotope experiment, that is, 49(m/z)/64(m/z)=0.82, is also found. In addition to the efficient formation of C$_2$H$_\text{n}$O (e.g., acetaldehyde, vinyl alcohol, ketene, and ethanol) directly supported by the IR spectra, the QMS data combined with the isotope-labeled experiment indicate that acetic acid may form as well and shows up via a desorption peak at 177 K with the highest mass signal of 60 (64) m/z in the experiment of C$_2$H$_2$ + O$_2$ + H (C$_2$H$_2$ + $^{18}$O$_2$ + H).

\subsection{Control experiments}
The surface chemistry of C$_2$H$_2$ + O$_2$ + H at 10 K clearly shows that the newly formed O-bearing products are  chemically connected, such as the isomers acetaldehyde and vinyl alcohol, as well as the sequential hydrogenation products: ketene, acetaldehyde, and ethanol. The isomer equilibrium between acetaldehyde and vinyl alcohol is a well-known example of keto-enol tautomerism; they are convertible and the reaction is generally in favor of acetaldehyde (keto-form) in the gas phase at standard conditions (room temperature and atmospheric pressure), as this contains the lowest chemical potential energy among the C$_2$H$_4$O isomers. Moreover, the interactions between the newly formed acetaldehyde (or vinyl alcohol) and impinging H-atoms are also expected in the experiments of C$_2$H$_2$ + O$_2$ + H to explain the formation of ketene and ethanol. Since pure vinyl alcohol is not commercially available, a set of control experiments starting from acetaldehyde was performed to (1) testify the unfavorable tautomerization from acetaldehyde to vinyl alcohol (i.e., clarifying the origin of vinyl alcohol) and (2) examine the possible formation route leading to the hydrogenated ethanol and the de-hydrogenated ketene. The experimental results are briefly presented here, and for the details of spectral identification the reader is referred to Appendix B.


In exp. c1, acetaldehyde ice was predeposited on the substrate at 10 K and then exposed to H-atoms (see Appendix B.1). The main experimental finding here is that acetaldehyde interacting with impinging H-atoms not only results in the formation of saturated alcohol (i.e., ethanol) at 10 K, which supports the theoretical conclusions by \cite{Charnley1997a} and confirms the laboratory findings by \cite{Bisschop2007}, but also leads to de-hydrogenation reactions forming ketene. Both products are firmly detected by RAIRs and TPD-QMS data. Similar forward and backward H-atom (or D-atom) reactions have experimentally been studied in much detail for another aldehyde (i.e., H$_2$CO) forming CH$_3$OH and CO, respectively, on grain surfaces \citep{Hidaka2009,Chuang2016,Minissale2016b}. There is no vinyl alcohol signal observed after H-atom addition reactions, indicating that the isomer rearrangement from acetaldehyde to vinyl alcohol triggered by impacting H-atoms is very unlikely because of the high activation barrier required ($\sim$282 kJ mol$^{-1}$, \citealt{Smith1991}).

\begin{figure}[]
	\begin{center}
		\includegraphics[width=90mm]{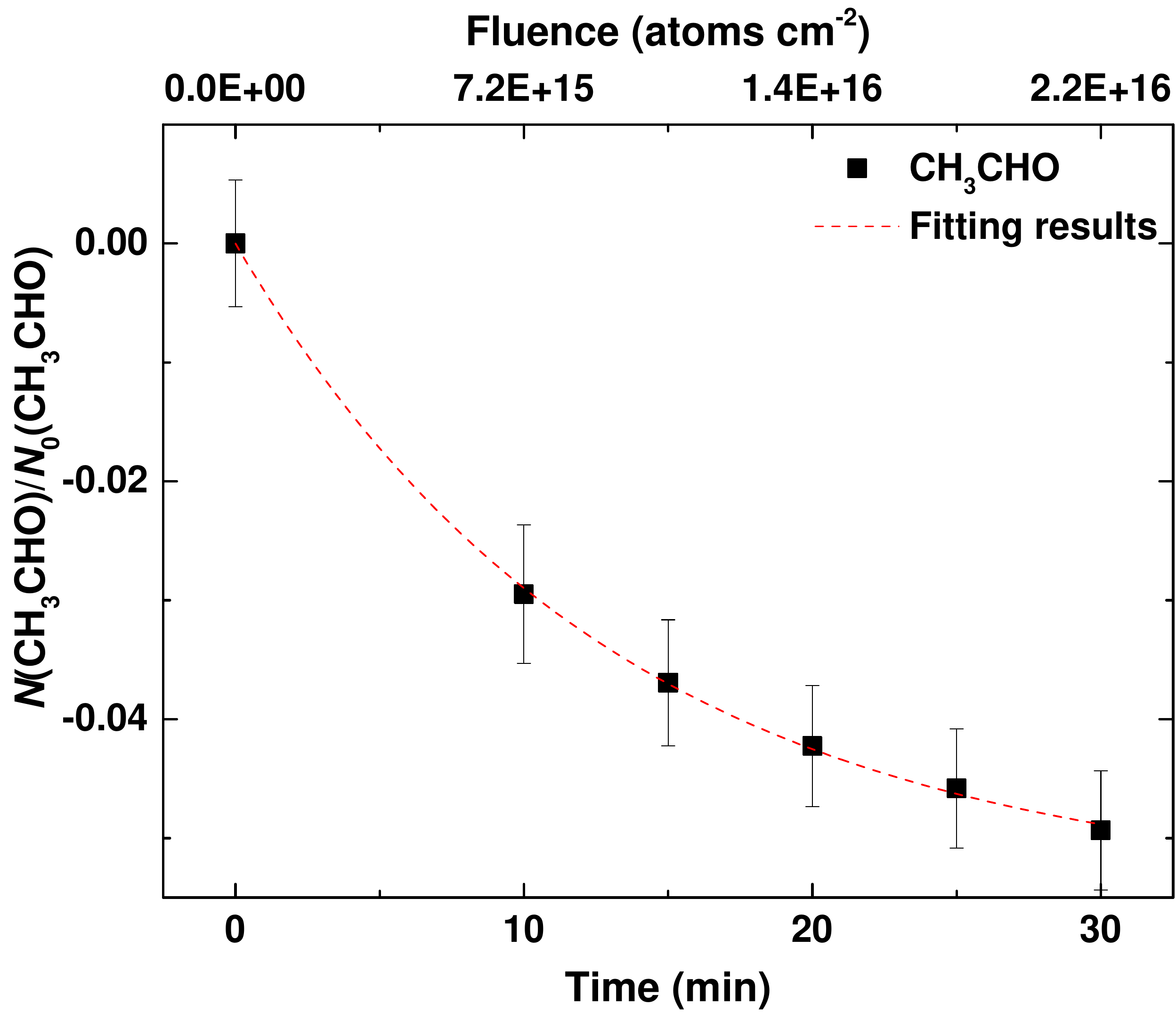}
		\caption{Kinetics of initial molecule (CH$_3$CHO) as a function of H-atom fluence. The red dashed line is presented for the fit results by using a first order kinetic rate equation.}
		\label{Fig6}
	\end{center}
\end{figure}

In Fig. \ref{Fig6}, a kinetic analysis of the relative abundance of the consumed acetaldehyde (i.e., $\triangle$\textit{N}(CH$_3$CHO)$\diagup$\textit{N}$_0$(CH$_3$CHO)) as a function of the duration of the H-atom addition experiment shows a single exponential decay as a function of H-atom fluence. As the abundance of H-atoms ([H]), which temporarily stay on the grain surface, is simply unknown, an “effective” rate constant, namely \textit{k}$_{\text{CH$_3$CHO+H}}$$\times$[H] in s$^{-1}$, is introduced here that is obtained by fitting the kinetics with a pseudo first-order kinetic rate equation, similar to that used in previous studies by \cite{Bisschop2007} and \cite{Kobayashi2017}. The derived effective rate constant of acetaldehyde hydrogenation at 10 K is $\sim$(1.3$\pm$0.1)$\times$10$^{-3}$ s$^{-1}$ for an H-atom exposure flux of 1.2$\times$10$^{13}$ atoms cm$^{-2}$s$^{-1}$ at 10 K. The reaction constant is in the same order of magnitude with that of CO + H (i.e., \textit{k}$_{\text{CO+H}}$$\times$[H]) obtained from a very similar study by \cite{Chuang2018a} and a few times higher than the value of \textit{k}$_{\text{CH$_3$CHO+H}}$$\times$[H] (i.e., $\sim$(3.8$\pm$0.8)$\times$10$^{-4}$ s$^{-1}$) reported by \cite{Bisschop2007} at $\sim$15 K. The discrepancy is most likely due to differences in experimental conditions, specifically ice temperature and H-atom exposure flux. 

The final product ratio of ethanol to ketene is $\sim$4.4$\pm$1.0, which is very similar to the product ratio (5.3$\pm$1.6) of methanol to carbon monoxide in a similar predeposition experiment of CO + H reported by \cite{Chuang2018a} at 10-14 K. These experimental findings show that the hydrogenated (forward) product is generally more abundant than the de-hydrogenated one.

The accumulated yield of ketene and ethanol in the predeposition experiment is limited by the H-atom penetration depth. Consequently, the resulting product ratio is strongly influenced by the successively accreting H-atoms and the formation of saturated species is favored as seen in this work and found in the literature \citep{Watanabe2002,Fuchs2009,Chuang2018a}. 

In order to directly compare the results of CH$_3$CHO + O$_2$ + H and C$_2$H$_2$ + O$_2$ + H codepositions, the CH$_3$CHO + H (exp. c2) and CH$_3$CHO + O$_2$ + H (exp. c3) codeposition experiments were performed for an identical O$_2$ to H-atoms ratio as in the C$_2$H$_2$ + O$_2$ + H codeposition experiment (see Appendix B.2).

In exp. c2, the codeposition of CH$_3$CHO + H shows stronger IR and QMS spectral signals of product ketene and ethanol than the data recorded from the predeposition experiment, in which the interactions are only limited to the upper layers of acetaldehyde. Vinyl alcohol is still absent in the simultaneous deposition experiment of CH$_3$CHO + H, which is in good agreement with previous experimental results implying an inefficient isomerization from keto to enol species. In addition to the detection of the (de-)hydrogenation products of acetaldehyde, the simultaneous deposition of CH$_3$CHO with O$_2$ and H-atoms were performed to study the formation of two-oxygen-bearing products formed through interactions of OH-radicals with acetaldehyde molecules (exp. c3). Acetic acid (CH$_3$COOH) is observed through its IR absorption features and QMS fragmentation pattern during the TPD experiment, further supporting the tentative identification in exps. 3 and 4 starting from C$_2$H$_2$ ice.

The control experiments presented here confirm that the surface chemistry of CH$_3$CHO with H-atoms and OH-radicals at 10 K efficiently leads to the formation of ketene, ethanol, and possibly acetic acid. The quantitative investigation in terms of product composition for all experiments and the possible formation mechanism are discussed in the next section. 

\section{Discussion}
\subsection{COMs composition}

\begin{figure}[t]
	\begin{center}
		\includegraphics[width=90mm]{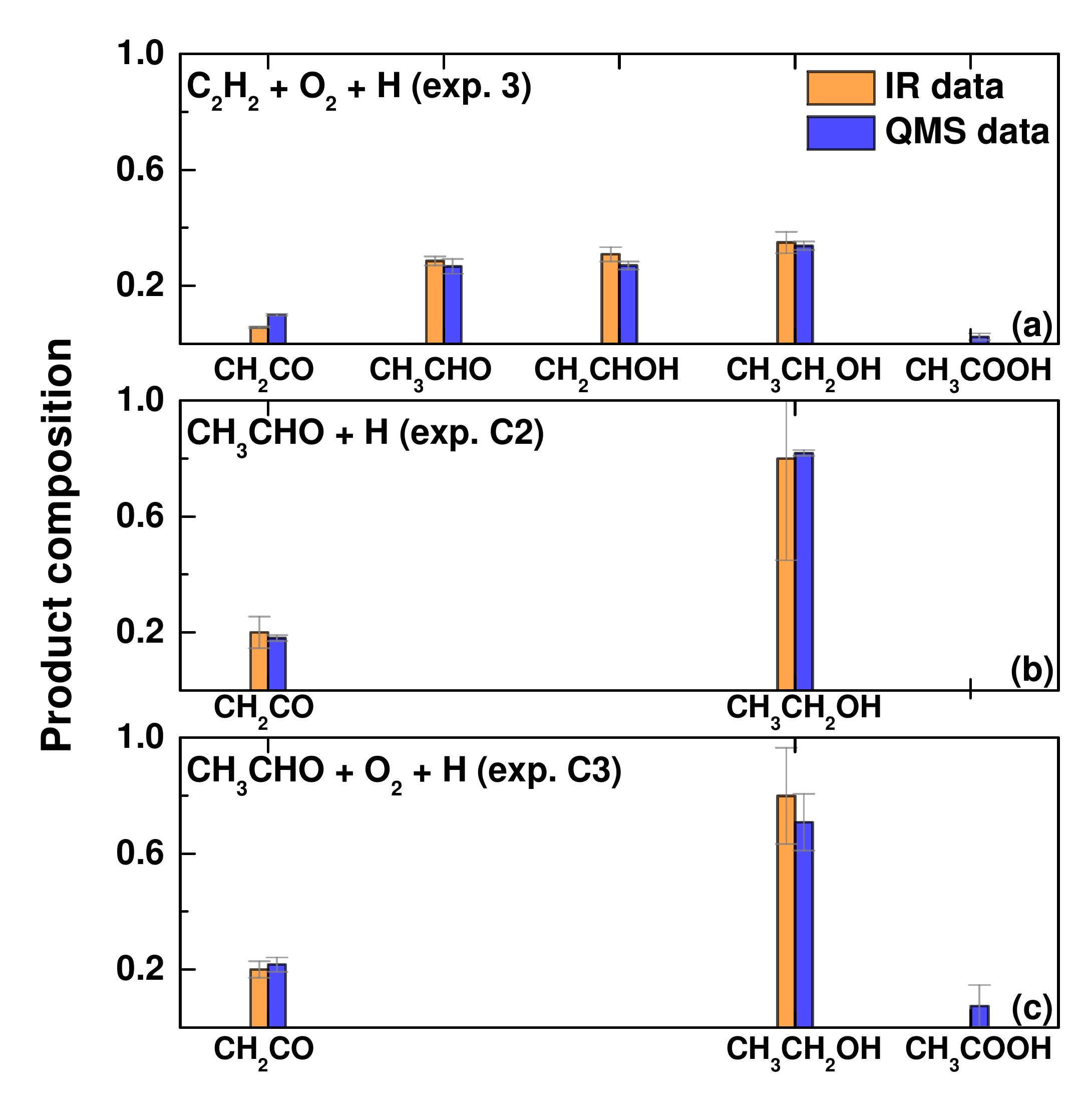}
		\caption{Relative product composition obtained in experiment (a) C$_2$H$_2$ + O$_2$ + H (exp. 3), (b) CH$_3$CHO + H (exp. c2), and (c) CH$_3$CHO + O$_2$ + H (exp. c3) by TPD-QMS and RAIRS, respectively.}
		\label{Fig10}
	\end{center}
\end{figure}

The formation yields or relative intensity of products have been derived by two independent methods as previously described in Section 2; via (1) RAIR data followed by the modified Beer-Lambert law using known or estimated absorption band strengths or via (2) TPD-QMS data followed by a series of calibrations taking into account any species’ fragmentation fraction, ionization cross-section, and detection sensitivity. The former one allows obtaining an absolute abundance (in column density), and the latter one, typically with better detection sensitivity, gives relative abundances of the newly formed products. In this work, both methods are utilized. The results are summarized in Fig. \ref{Fig10}.

Figure \ref{Fig10} shows the product composition derived from experiments C$_2$H$_2$ + O$_2$ + H (exp. 3; upper panel (a)), CH$_3$CHO + H (exp. c2; middle panel (b)), and CH$_3$CHO + O$_2$ + H (exp. c3; bottom panel (c)), respectively. In exp. 3, ketene, acetaldehyde, vinyl alcohol, and ethanol are all together observed, both in the IR and QMS data, and traces of acetic acid can only be quantified from mass spectra only. The resulting relative composition amounts to: 0.10$\pm$0.01  for ketene, 0.27$\pm$0.03 for acetaldehyde, 0.27$\pm$0.01  for vinyl alcohol, 0.34$\pm$0.02  for ethanol, and 0.02$\pm$0.01 for acetic acid, respectively. The derived composition from the RAIR data shows a very similar trend, namely 0.06$\pm$0.01, 0.28$\pm$0.02, 0.31$\pm$0.03, and 0.35$\pm$0.04 for ketene, acetaldehyde, vinyl alcohol, and ethanol, respectively. The small discrepancy between these two methods is most likely due to different uncertainties in the applied conversion factors, such as IR absorption band strength and QMS ionization cross-section, but the overall findings are rather consistent. Under our experimental conditions, for example, ethanol is the most abundant product and ketene only has less than 10\% of the total yield. It indicates that H-atom additions are more steering than H-abstraction reactions, effectively favoring the formation of the fully saturated species on the surfaces. Both the QMS and IR derived composition of vinyl alcohol are approximately the same within the reported errors suggesting that vinyl alcohol, in the present work, is exclusively formed at low temperature through atom and radical addition reactions to C$_2$H$_2$; the derived final composition is constant before and after TPD experiments. The QMS and IR abundance ratios of vinyl alcohol with respect to the sum of the other COMs (acetaldehyde-related species), namely \textit{N}(CH$_2$CHOH)$\diagup$\textit{N}(CH$_3$CHO + CH$_2$CO + CH$_3$CH$_2$OH + CH$_3$COOH) are $\sim$0.4 and $\sim$0.4, respectively. The ratio suggests that chemical reactions starting from vinyl alcohol are very efficient forming other products. The similar values reported by IR and QMS also confirm that these products are most likely formed at 10 K, rather than induced by thermal processing. 

In exp. c2, only ketene and ethanol are detected, verifying the chemical links between ketene, acetaldehyde and ethanol. It should be noted that vinyl alcohol is not observed in IR and QMS data at all, implying that tautomerization from acetaldehyde to vinyl alcohol (i.e., enolization) is unfavorable in a temperature range from 10 to 300 K. The corresponding composition fractions from TPD-QMS data are 0.18$\pm$0.01 and 0.82$\pm$0.01 for ketene and ethanol, respectively. Similar results can also be derived from the IR data showing  0.20$\pm$0.05 for ketene and  0.80$\pm$0.35 for ethanol. Furthermore, in exp. c3, ketene and ethanol are still the two dominant species with traces of acetic acid. TPD-QMS (and IR) data shows composition fractions of  0.22$\pm$0.02 (0.20$\pm$0.03), 0.71$\pm$0.10 (0.80$\pm$0.17), and  0.07$\pm$0.07 for ketene, ethanol, and acetic acid, respectively. Vinyl alcohol is not observed, confirming the previous conclusion that vinyl alcohol can only be formed through the reactions between C$_2$H$_2$ and OH-radicals as well as H-atoms, rather than converting from the preformed acetaldehyde.

Ketene and ethanol are two newly formed products that are commonly detected in both experiments starting from C$_2$H$_2$ and from CH$_3$CHO. 
Under similar experimental conditions, that is, a constant ratio of H-atoms to O$_2$ in C$_2$H$_2$ + O$_2$ + H (exp. 3) and CH$_3$CHO + O$_2$ + H (exp. c3), a similar ratio of ketene to ethanol is observed by QMS; 0.30$\pm$0.01 and 0.31$\pm$0.03 for exp. 3 and exp. c3, respectively. The experimental findings suggest that these two products are chemically correlated and share the same formation mechanism. However, the derived ratio of ketene to ethanol (i.e., 0.22$\pm$0.01) in experiment CH$_3$CHO + H (exp. c2) is $\sim$25\% less than the ratio found in exp. 3 and exp. c3. It matches the expectation that more H-atoms are available to hydrogenate species into the saturated product (e.g., ethanol). This finding also follows from the hydrogenation experiment of predeposited CH$_3$CHO (exp. c1) as products and parent species reach a steady state; given a successive H-atom accretion on the initial CH$_3$CHO ices, the product ratio of ketene to ethanol from QMS data is even smaller (i.e., 0.11$\pm$0.03). Our experimental results clearly demonstrate that starting with different parent species (C$_2$H$_2$ or CH$_3$CHO) under different experimental conditions, such as H-atoms only versus H-atoms and OH-radicals, leads to different final products and composition, as one may expect.


\subsection{COM formation mechanisms}

As described in the previous work by \cite{Hiraoka2000} and \cite{Kobayashi2017}, the interaction between C$_2$H$_2$ and H-atoms shows efficient production of C$_2$H$_4$ and C$_2$H$_6$ through successive H-atom addition reactions;
\begin{equation}\label{Eq3}
\text{C$_2$H$_2$}\xrightarrow{\text{+H}}\text{C$_2$H$_3$}\xrightarrow{\text{+H}}\text{C$_2$H$_4$}\xrightarrow{\text{+H}}\text{C$_2$H$_5$}\xrightarrow{\text{+H}}\text{C$_2$H$_6$}.
\end{equation}

From ab initio theoretical studies, the reported activation barrier of C$_2$H$_2$ + H (i.e., 17.99 kJ mol$^{-1}$; ~2170 K) is higher than that of C$_2$H$_4$ + H (i.e., 11.72 kJ mol$^{-1}$; ~1428 K) \citep{Miller2004,Kobayashi2017}. These calculations greatly explain the experimental results showing that the final production of C$_2$H$_6$ is more abundant than the C$_2$H$_4$ yield.
 
In experiments involving O$_2$ molecules and H-atoms, C$_2$H$_2$ hydrogenation is not the only H-addition reaction channel. The reactions between H-atoms and O$_2$ molecules mainly result in OH-radical formation through a series of barrierless (or at least negligible barrier) reactions;
\begin{equation}\label{Eq4}
\text{O$_2$}\xrightarrow{\text{+H}}\text{HO$_2$}\xrightarrow{\text{+H}}\text{2OH}.
\end{equation}
The last reaction step could also lead to H$_2$O$_2$ and H$_2$O \citep{Cuppen2010}. 

One would expect that OH-radicals are abundantly formed on grain surfaces through either reaction \ref{Eq4} or converting available O-atoms (as aforementioned in the Introduction) by associating with abundant H-atoms applied in our experiments. Therefore, the widely available OH-radicals are expected to trigger further surface chemistry, such as radical-atom(radical) and radical-molecule reactions. 
Moreover, the theoretical calculations and experimental studies consistently show that the activation barrier of C$_2$H$_2$ + OH is in the range of 5.4-8.0 kJ mol$^{-1}$ (654-956 K), which is a few times less than the required barrier of C$_2$H$_2$ hydrogenation in the gas phase \citep{Michael1979,Smith1984,Senosiain2005,McKee2007}. \cite{Basiuk2004} report that the OH-radical addition reaction to C$_2$H$_2$ has no transition state. All these findings imply that an OH-radical is also very likely to react with the initial C$_2$H$_2$ through a radical-molecule reaction, once it is formed adjacently to the reactants in the solid state (i.e., without the need of diffusion) through reaction;
\begin{equation}\label{Eq5}
\text{C$_2$H$_2$ + OH}\rightarrow\text{CHCHOH}.
\end{equation}

The unstable hydroxyvinyl radical (CHCHOH) could then react with H-atoms forming vinyl alcohol (CH$_2$CHOH) via the barrierless reaction:
\begin{equation}\label{Eq6}
\text{CHCHOH + H}\rightarrow\text{CH$_2$CHOH}.
\end{equation}  

Both reaction \ref{Eq5} and \ref{Eq6} are exothermic, namely $\sim$132.2 and ~464.0 kJ mol$^{-1}$, respectively \citep{Basiuk2004}. It is important to note that the C$_2$H$_2$ hydrogenation scheme in reaction \ref{Eq3} is still an important reaction channel due to relatively abundant and mobile H-atoms applied in this work. Therefore, the OH-radical recombination with the intermediate radicals, such as C$_2$H$_3$ or C$_2$H$_5$ formed in reaction \ref{Eq3} provides another possible channel for the formation of vinyl alcohol via:
\begin{subequations}
\begin{equation}\label{Eq7a}
\text{C$_2$H$_3$ + OH}\rightarrow\text{CH$_2$CHOH}
\end{equation}
or for ethanol via:
\begin{equation}\label{Eq7b}
\text{C$_2$H$_5$ + OH}\rightarrow\text{CH$_3$CH$_2$OH}.
\end{equation}
\end{subequations}  

The contribution of the above reactions \ref{Eq7a} and \ref{Eq7b} is expected to be minor due to the requirement of two adjacent radicals and the relatively high reactivity of radicals reacting with abundant H-atoms via reaction \ref{Eq3}. 
Alternatively, the barrierless reactions between OH-radicals and  C$_2$H$_4$ were reported in laboratory and theoretical studies in the gas phase by \cite{Zellner1984} and \cite{Cleary2006};
\begin{equation}\label{Eq8}
\text{C$_2$H$_4$ + OH}\rightarrow\text{CH$_2$CH$_2$OH}.
\end{equation}
The associated product can subsequently recombine with an H-atom forming ethanol:
\begin{equation}\label{Eq9}
\text{CH$_2$CH$_2$OH + H}\rightarrow\text{CH$_3$CH$_2$OH}.
\end{equation}
A unimolecular isomerization from CH$_2$CH$_2$OH to CH$_3$CH$_2$O is less likely due to the extra energy required to overcome the rearrangement barrier, which is 13.6 kJ mol$^{-1}$ higher than the initial potential energy of reactants in the gas phase \citep{Sosa1987}.

\begin{figure*}[t]
	\vspace{-2cm}
	\begin{center}
		\includegraphics[width=100mm]{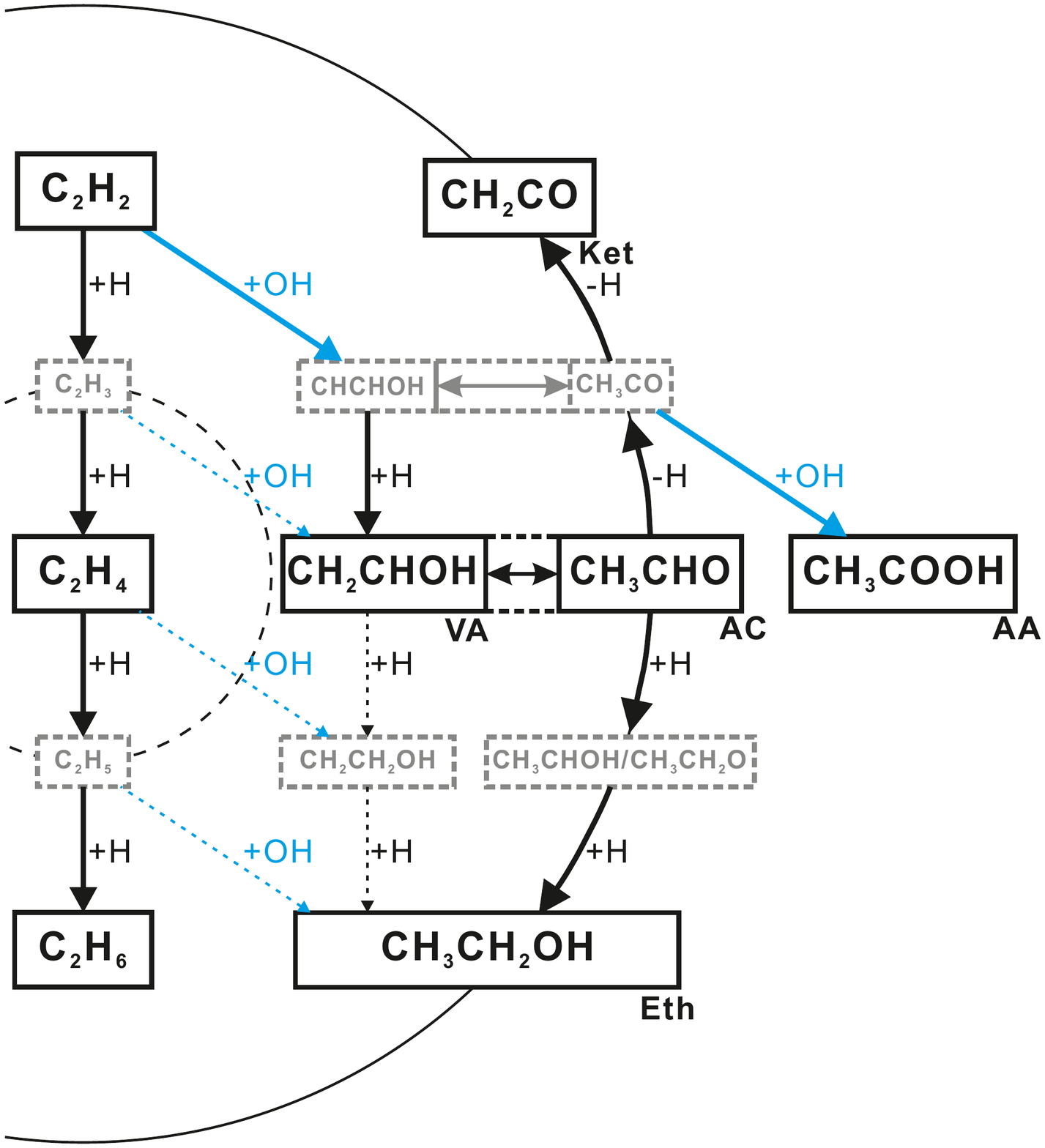}
		\vspace{-2cm}
		\caption{Proposed solid-state reaction scheme of C$_2$H$_2$interacting with H-atoms and OH-0radicals. Solid arrows indicate reaction pathways confirmed or suggested in this work; black lines represent H-atom reactions and blue lines represent OH-radical addition reactions. Dashed arrows indicate other possible reaction pathways which could also lead to the investigated COMs. VA: vinyl alcohol; AC: acetaldehyde; Ket: ketene; Eth: ethanol; AA: acetic acid. }
		\label{Fig12}
		
	\end{center}
	
\end{figure*} 

On the basis of the work presented here, we cannot exclude these additional reactions \ref{Eq7a}, \ref{Eq7b}, \ref{Eq8}, and \ref{Eq9}. However, these additional formation channels of vinyl alcohol and ethanol would require one or two more H-atom addition steps of C$_2$H$_2$ for reaction \ref{Eq7a} and \ref{Eq8}, respectively, prior to reacting with OH-radicals. H-abstraction reactions induced by OH-radicals on stable hydrocarbons are generally unfavored due to an endothermic reaction of C$_2$H$_2$ + OH$\rightarrow$C$_2$H + H$_2$O and a high activation barrier (i.e., 30.6 kJ mol$^{-1}$) required in C$_2$H$_4$ + OH$\rightarrow$C$_2$H$_3$ + H$_2$O \citep{Smith1984,Cleary2006}. 

The newly formed vinyl alcohol, most likely through reactions \ref{Eq5} and \ref{Eq6} could subsequently be isomerized to acetaldehyde, which contains the lowest potential energy (i.e., 62.3 kJ mol$^{-1}$ lower than vinyl alcohol), with the help of the excess energy obtained from the (highly) exothermic association reactions \ref{Eq5} and \ref{Eq6}; 
\begin{equation}\label{Eq10}
\text{CH$_2$CHOH (enol-form)}\leftrightarrow\text{CH$_3$CHO (keto-form)}.
\end{equation}

In the gas phase, the isomerization from vinyl alcohol to acetaldehyde requires a relatively high activation barrier of $\sim$250 kJ mol$^{-1}$ through an “intramolecular” pathway; unimolecular rearrangement via hydrogen transfer within the molecule (illustrated in figure 1 in \citealt{Apeloig1990}). However, this value can be dramatically reduced through an “intermolecular” channel; H-atoms exchange (double-hydrogen-shift; DHS) with surrounding molecules (illustrated in figure 4 in \citealt{Apeloig1990}). Theoretical studies in the gas phase suggest that multiple molecules, such as H$_2$O, C$_2$H$_2$, and acids, can catalyze the isomerization \citep{Klopman1979,daSilva2010}. Moreover, in the liquid phase, the water chain (H$_2$O)$_\text{n}$ intervention in the ketonization of vinyl alcohol was theoretically and experimentally investigated showing a significant drop in potential energy down to 91.2 and 49.5 kJ mol$^{-1}$, respectively \citep{Capon1982,Lledos1986}. In this work, the simultaneous detection of abundant vinyl alcohol and acetaldehyde supports an efficient ketonization in the solid state for the investigated experimental conditions.  We cannot exclude the possibility of the similar (enol-keto) isomerization mechanism between surrounding H$_2$O (or H$_2$O$_2$) and hydroxyvinyl radical from reaction \ref{Eq5};
\begin{equation}\label{Eq11}
\text{CHCHOH (enol-radical)}\leftrightarrow\text{CH$_2$CHO (keto-radical)}.
\end{equation}

However, the above reaction route has to compete with the (efficient) recombination reaction with H-atoms, namely reaction \ref{Eq6}, leading to vinyl alcohol. More experimental and theoretical studies are required to examine the validity of keto-enol isomerization in radical analogs. The obtained product ratio of enol and keto is constant before and after TPD experiment, suggesting that the equilibrium of isomerization is immediately determined once the species synthesis through reactions \ref{Eq6} and \ref{Eq10} (or reactions \ref{Eq5} and \ref{Eq11}).
 
The absence of ethylene oxide is consistent with the expectation that ethylene oxide contains the highest potential energy among the three C$_2$H$_4$O isomers; spontaneous rearrangement is very unlikely. The nondetection result also verifies an O-free environment in our experiments, because the reaction of C$_2$H$_4$ + O primarily leads to the formation of ethylene oxide in the solid state with an very low activation energy of $\sim$190 K \citep{Ward2011}. In space, determining which isomers are primarily formed on grain surface is done by the competition between OH-radical and O-atom addition reactions to hydrocarbons. However, O-atoms are expected to be efficiently converted into OH-radicals or H$_2$O due to abundant H-atoms available in molecular clouds.

The resulting acetaldehyde interacts with H-atoms forming second-generation products through reactions;
\begin{subequations}
	\begin{equation}\label{Eq12a}
	\text{CH$_3$CHO}\xrightarrow{\text{+H}}\text{CH$_3$CH$_2$O/CH$_3$CHOH}\xrightarrow{\text{+H}}\text{CH$_3$CH$_2$OH}
	\end{equation}
	\begin{equation}\label{Eq12b}
	\text{CH$_3$CHO}\xrightarrow{\text{+H(-H$_2$)}}\text{CH$_3$CO/CH$_2$CHO }\xrightarrow{\text{+H(-H$_2$)}}\text{CH$_2$CO}.
	\end{equation}
\end{subequations} 

The reaction \ref{Eq12a} has been experimentally studied by \cite{Bisschop2007}, and the H-abstraction (dehydrogenation) in reaction \ref{Eq12b} has been subject of numerous experimental studies in the gas phase (\citealt{Ohmori1990} and reference therein). It is noted that the newly formed CH$_2$CO is able to be further hydrogenated forming acetaldehyde again with a small barrier reported in the gas phase \citep{Umemoto1984} or associated with OH-radical forming CH$_2$COOH, which is similar to the currently studied hydroxylation of unsaturated hydrocarbon. Similar forward (addition) and backward (abstraction) findings have also been reported in CO-H$_2$CO-CH$_3$OH reaction scheme in the solid state \citep{Hidaka2007,Chuang2016,Minissale2016b}. Additionally, the H-abstraction reaction through CH$_3$CHO + OH $\rightarrow$ CH$_3$CO + O has also been proposed to be a barrierless reaction in the gas phase \citep{Taylor2006}. The intermediate radicals are typically undetectable due to their high reactivity compared to stable molecules. These reactive radicals are not only consumed by H-atoms, but also recombine with OH-radicals through the reactions:
\begin{subequations}
	\begin{equation}\label{Eq13a}
	\text{CH$_3$CO + OH}\rightarrow\text{CH$_3$COOH}
	\end{equation}
	\begin{equation}\label{Eq13b}
	\text{CH$_2$CHO + OH}\rightarrow\text{HCOCH$_2$OH}.
	\end{equation}
\end{subequations} 

In experiments C$_2$H$_2$ + O$_2$ + H (exp. 3) and CH$_3$CHO + O$_2$  + H (exp. c3), acetic acid is the only detected product consisting of two O-atoms. This observation implies that the H-abstraction of acetaldehyde likely takes place on the -CHO end resulting in a CH$_3$CO radical, which is then available to react with an OH-radical forming acetic acid through reaction \ref{Eq13a}, rather than forming glycolaldehyde via reaction \ref{Eq13b}.

The proposed reaction network of COMs based on all investigated reaction routes is summarized in Fig. \ref{Fig12} starting from the interactions between C$_2$H$_2$ and H-atoms as well as OH-radicals. As confirmed in this work, from top to down, the successive hydrogenation reactions of C$_2$H$_2$ efficiently lead to C$_2$H$_4$ and C$_2$H$_6$. The competition between H-atom and OH-radical additions to C$_2$H$_2$ may lead to the production of CHCHOH radicals and then vinyl alcohol (VA) after associating with H-atoms. An efficient isomerization from vinyl alcohol to acetaldehyde (AC) probably through the so-called intermolecular pathways is observed. The H-atom addition and abstraction reactions further convert acetaldehyde into ketene (Ket) and ethanol (Eth). Simultaneous forward and backward reactions very likely enhance the lifetime of intermediate radicals, such as CH$_3$CO on grain surfaces, which could have a higher possibility of associating with additional OH-radical forming heavier COMs, such as acetic acid (AA).

\section{Astrochemical implication and conclusions}
This experimental study proves the solid-state formation of COMs described by the formula C$_2$H$_\text{n}$O, such as acetaldehyde, vinyl alcohol, ketene, and ethanol, but not ethylene oxide, through the interactions of C$_2$H$_2$ with H-atoms and OH-radicals. These chemically correlated COMs are firmly identified all together in the interstellar ice analogs along with an H$_2$O-ice forming scenario. In the early stage of star evolution, H-atoms and OH-radicals are expected to be the main chemical triggers that can lead to hydrogen-saturated molecules through surface radical-radical reactions, such as H + H$\rightarrow$ H$_2$, N + 3H$\rightarrow$ NH$_3$, C + 4H$\rightarrow$ CH$_4$, and H + OH$\rightarrow$H$_2$O, or even radical-molecule reactions (see review paper: \citealt{Hama2013,Linnartz2015}). In this work, we extend the existing surface radical-molecule reaction network by including simple hydrocarbon molecules (e.g., C$_2$H$_2$), which have been proposed to be necessary precursor species for various COMs \citep{Basiuk2004,Charnley2004}. In addition to C$_2$H$_2$ hydrogenation channels, the experimental results show the validity of direct OH-radical addition reactions (hydroxylation) to C$_2$H$_2$ that open an efficient formation pathway leading to complex species without additional triggers by energetic processes in molecular clouds.

The laboratory study presented here introduces an additional COM forming mechanism through OH-radical additions to hydrocarbons before CO catastrophically freezes out. The investigated COM reaction network is consistent with recent astronomical observations toward cold and dark clouds in star-forming regions commonly showing very abundant acetaldehyde, namely $\sim$10$^{-10}$-10$^{-9}$ with respect to \textit{n}(H$_2$), at low temperature \citep{Bacmann2012,Cernicharo2012,Taquet2017}. Furthermore, the detections of acetaldehyde in conjunction with ketene in translucent clouds are well supported by our experimental findings indicating that these C$_2$H$_\text{n}$O species could have a similar origin before CO molecules accrete and then participate in the COM synthesis on grain surfaces, typically leading to different classes of C$_2$H$_\text{n}$O$_2$ species.
 
Infrared observations of interstellar dust in diffuse clouds commonly show the saturated aliphatic features around 3.4 $\mu$m, suggesting the existence of carbonaceous dust that is attached by various hydrocarbon functional groups \citep{Sandford1991,Mennella2003}. However, these hydrocarbon features are typically absent in dense molecular clouds, and the elimination mechanism is still under debate \citep{Brooke1995,Chiar1996,Godard2011}. One of the possible explanations is that a rapid cleavage of C-H or C-C bonds induced by high energetic photons or cosmic particles takes place when cloud density and A$_v$ are still low \citep{Pendleton1999,Mate2016}. Subsequent de-hydrogenation processing could further lead to the closure of dissociated fragments (i.e., through two unpaired electrons recombine), forming stable hydrocarbon molecules bound by hybridized bonds, such as \textit{sp}$^2$ (i.e., C$=$C) or \textit{sp}$^1$ (i.e., C$\equiv$C) \citep{Dartois2017}. Later, as cloud density increases, these newly formed species may react with impacting H-atoms forming completely saturated carbon chains or interact with the surrounding OH-radicals resulting in aldehydes or alcohols through surface nonenergetic processing. It is important to note that the solid-state OH-radicals are mainly formed on grain surfaces via the reactions of O + H or via photon or cosmic ray triggered H$_2$O dissociation leading to OH, rather than directly accrete from the gas phase due to a low OH-abundance in clouds (the gaseous ratio of OH-radicals to H-atoms is $\sim$10$^{-7}$ \citep{Ohishi1992,Roueff1996}. Since H$_2$O is the major constituent of interstellar ices and primarily formed through the solid-state reactions involving OH-radicals; OH + H$\rightarrow$H$_2$O, OH + OH$\rightarrow$H$_2$O + O, and OH + H$_2$$\rightarrow$H$_2$O + H, OH-radicals are expected to abundantly be present on grain surfaces as shown by the chemical simulation by \citealt{Cuppen2007} (i.e., the ratio of OH/H$_2$O in the range of 0.2-1.0 under translucent cloud conditions), and to react with hydrocarbons. Moreover, the dominant H$_2$O ice may play a catalytic role in the keto-enol isomerization enhancing the aldehyde formation through intermolecule pathways. Once A$_v$$\geq$10 (i.e., \textit{n}(H)$\approx$10$^{4-5}$ cm$^{-3}$), the gaseous CO catastrophically freezes out onto grain surfaces and becomes the dominant constituent in the ice chemistry. It results in a dramatic depletion of OH-radicals because of abundant CO species (\textit{N}(CO)/\textit{N}(H)$\approx$1-10) that not only compete for available H-atoms, but also consume OH-radicals forming CO$_2$ on grain surfaces through de-hydrogenation of the HO-CO radical. \citep{Ioppolo2011}. The availability of OH-radicals controls the efficiency of the studied hydrocarbon hydroxylation. However, the intermediate products (e.g., HCO) along with the CO hydrogenation scheme may provide alternative pathways to form COMs; recombining with hydrocarbon fragments (e.g., C$_2$H$_3$) leads to formation of propanal and propanol reported by \cite{Qasim2019b} or associating with other CO-bearing radicals (e.g., CH$_2$OH and CH$_3$O) results in sugar relatives \citep{Fedoseev2015b,Chuang2016}. It should be noted that the present and these latter studies focus on different evolutionary stages, and that reactions may result in similar reaction products, but following different reaction pathways.

The proposed COM formation route through hydrocarbon and OH-radicals is not limited to simple species (e.g., C$_2$H$_2$). The OH-radical addition reactions may also be effective on larger hydrocarbons, such as H-(C$\equiv$C)$_\text{n}$-H and H$_3$C-(C$\equiv$C)$_\text{n}$-H (n$>$1), forming biologically relevant molecules. This provides a potential route to form large amphiphilic molecules (i.e., a chemical compound possessing both hydrophilic and lipophilic properties), with a potential link to astrobiology and formation of vesicles (i.e., an intracellular membrane). The efficient reactivity of OH-radicals with propyne (HCCCH$_3$) was recently demonstrated by \cite{Qasim2019a} and results in the formation of propenols and propanols. It is noted that the isomerization (i.e., the conversion between propenol and propanal) was not reported, probably due to water-poor experimental conditions in contrast to the experiments presented here.

In addition to acetaldehyde, ethylene oxide, one of its isomers, has recently been detected in the prestellar core L1689B with a small fractional abundance of $\chi$$_\text{H$_2$}$$\approx$3.5$\times$10$^{-12}$, which is $\sim$30 times less than acetaldehyde \citep{Bacmann2019}. The formation mechanism of ethylene oxide has been proposed through interactions between C$_2$H$_4$ and thermal O-atoms or suprathermal O-atoms originating from energetic dissociation of simple CO$_2$. For example, \cite{Ward2011} showed that the primary product in the thermal O-atom addition reactions to C$_2$H$_4$ is ethylene oxide, and the reactions between suprathermal O-atoms and C$_2$H$_4$ result in a comparable amount of ethylene oxide as acetaldehyde \citep{Bennett2005b,Bergner2019}. However, observational data from interstellar clouds to protostars all report that acetaldehyde is more abundant than other isomers (i.e., vinyl alcohol and ethylene oxide). It implies that a dominant formation mechanism leading to acetaldehyde must take place in the early molecular clouds, such as solid state radical chemistry. Our experimental results support the idea that these COMs form through the reactions of simple C$_2$H$_2$, which originates from PAH$^+$ photodissociation or HAC erosion, with H-atoms and OH-radicals on grain surfaces, as summarized in Fig. \ref{Fig12}. For our experimental conditions (with a relatively short laboratory time-scale), the isomeric ratio of vinyl alcohol to acetaldehyde has been derived upon formation at 10 K. However, one would expect for astronomical time-scales that the final gas-phase equilibrium between these two isomers would be dominated by acetaldehyde as observational data suggested in prestellar cores (i.e., the upper limit ratio of \textit{N}(CH$_3$CHO)$\diagup$\textit{N}(CH$_2$CHOH)>9.2, \citealt{Bacmann2019}).\\

Below, the main experimental findings of this work investigating the surface chemistry of simple hydrocarbon C$_2$H$_2$ with H-atoms and OH-radicals (without CO) at 10 K are summarized.

   \begin{enumerate}
	\item	In addition to the efficient hydrogenation of C$_2$H$_2$ forming C$_2$H$_4$ and C$_2$H$_6$, C$_2$H$_2$ can serve as a starting point for the formation of complex organic molecules through a series of OH-radical and H-atom addition reactions along with the simultaneous H$_2$O (or H$_2$O$_2$) formation under molecular cloud conditions; C$_2$H$_2$$\xrightarrow{\text{+OH}}$CHCHOH$\xrightarrow{\text{+H}}$CH$_2$CHOH$\diagup$CH$_3$CHO $\xrightarrow{\text{+2H}}$CH$_3$CH$_2$OH (see Fig. 7).

	\item Both TPD-QMS and IR analyses show a similar product composition in the experiment of C$_2$H$_2$ + O$_2$ + H.  The obtained ratios of ketene: acetaldehyde: vinyl alcohol: ethanol: acetic acid are 0.10: 0.27: 0.27: 0.34: 0.02 and 0.06: 0.28: 0.31: 0.35: $\sim$0 for QMS and IR, respectively. This implies that there is no extra COM formation during thermal processing.
	
    \item The simultaneous detection of vinyl alcohol and acetaldehyde implies that the isomerization from vinyl alcohol to acetaldehyde occurs upon formation of vinyl alcohol or vinyl alcohol radical, and likely proceeds through intermolecular H-atom exchange with the surrounding H-bearing species in the solid state at 10 K. However, there is no evidence of the reversed mechanism from acetaldehyde to vinyl alcohol during H-atoms and OH-radicals exposure or upon thermal processing. The equilibrium point between these two isomers is expected to shift to acetaldehyde on a molecular cloud time-scale.
    
    \item The successive hydrogenation of predeposited acetaldehyde ice at 10 K leads to the formation of ketene through H-atom abstraction reactions and ethanol through H-atom addition reactions. The derived effective rate constant (\textit{k}$_{\text{CH$_3$CHO+H}}$$\times$[H]) of acetaldehyde hydrogenation is $\sim$(1.3$\pm$0.1)$\times$10$^{-3}$ s$^{-1}$ for the experimental conditions and the utilized H-atom flux applied here. The obtained ratio of ketene to ethanol in predeposition and codeposition experiments favors the saturated product (i.e., 0.11-0.30) and depends on the initial ice composition and available H-atom concentration on grain surfaces.
   \end{enumerate}

In the recent past, the formation of acetaldehyde and its derivatives in the solid state has extensively been studied through energetic processes, such as UV-photons, electrons, and protons, on preexisted interstellar ices \citep{Hudson2003,Bennett2005a,Bennett2005b,Oberg2009a,Bergner2019}. These previous studies show the formation of acetaldehyde through either the recombination of dissociation fragments or the interactions between (hot) O-atoms and hydrocarbons, such as C$_2$H$_4$ and C$_2$H$_6$. The proposed formation pathways of COMs derived from the experiments presented here further complements the solid state reaction network leading to molecular complexity in the ISM, that is, it does not conflict with other energetic mechanisms already proposed and even does not exclude possible gas phase routes. In turn, which of these formation mechanisms dominates strongly depends on the physical and chemical conditions in star-forming regions. It needs future laboratory studies and astrochemical simulations to proceed in order to derive the full chemical picture. The routes identified here are argued to play a major role under translucent cloud conditions.

\begin{acknowledgements}
This research was funded through a VICI grant of NWO (the Netherlands Organization for Scientific Research), an A-ERC grant 291141 CHEMPLAN, and has been performed within the framework of the Dutch Astrochemistry Network. Financial support by NOVA (the Netherlands Research School for Astronomy). G.F. acknowledges the financial support from the European Union’s Horizon 2020 research and innovation program under the Marie Sklodowska-Curie grant agreement no. 664931. S.I. acknowledges the Royal Society for financial support and the Holland Research School for Molecular Chemistry (HRSMC) for a travel grant. T.H. acknowledges support from the European Research Council under the Horizon 2020 Framework Program via the ERC Advanced Grant Origins 83 24 28. This work has been supported by the project PRIN-INAF 2016 The Cradle of Life - GENESIS-SKA (General Conditions in Early Planetary Systems for the rise of life with SKA).
\end{acknowledgements}

\bibliography{Ref}{}
\bibliographystyle{aa}

\begin{appendix}
\section{IR spectrum deconvolution of acetaldehyde and vinyl alcohol}
\begin{figure}[b]
	\begin{center}
		\includegraphics[width=90mm]{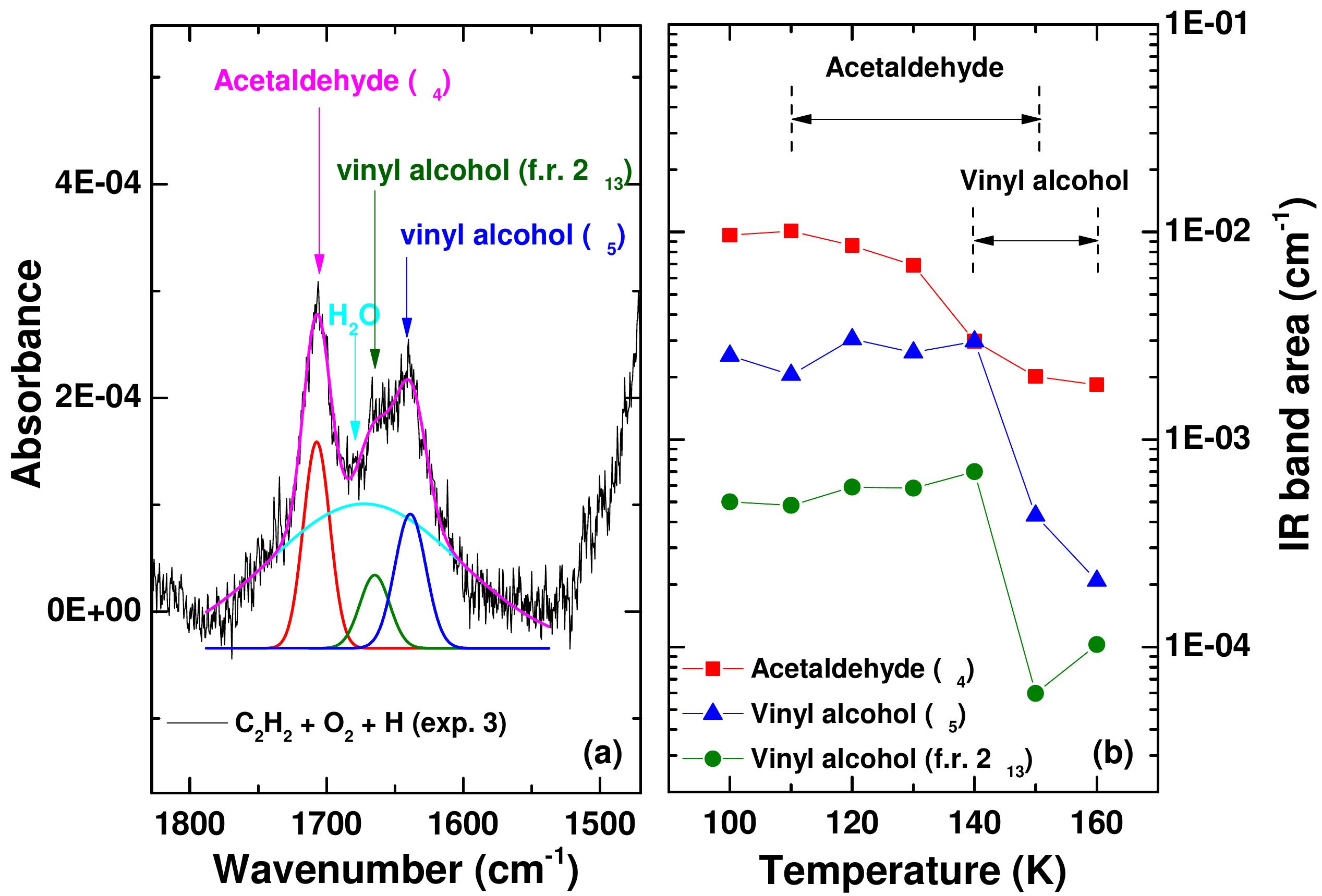}
		\caption{
			(a): IR features fit by Gaussian functions for the peaks of acetaldehyde, water, and vinyl alcohol in the C$_2$H$_2$ + O$_2$ + H study (exp. 3). (b): Integrated area of the IR figures as a function of temperature; a decrease reflects thermal desorption during TPD experiment. 	
		}
		\label{Fig.A1}
	\end{center}
\end{figure}

Figure \ref{Fig.A1}(a) presents an example of IR feature fits by Gaussian functions for the newly formed products, that is, acetaldehyde (peak on the left), water (broad profile), and vinyl alcohol (two peaks on the right), in the IR range from 1800 to 1550 cm$^{-1}$. The evolution of absorption area for the corresponding IR peaks during TPD experiments is shown in Fig. \ref{Fig.A1}(b). The peak of acetaldehyde (i.e., $\nu$$_4$) starts desorbing at 110 K, and the two peaks originating from vinyl alcohol (i.e., $\nu$$_5$ and fermi resonance with 2$\nu$$_{13}$) simultaneously deplete at $\sim$140 K. The correlation between temperatures for which the IR signals decrease and the QMS data increase correlate well. These results further support the IR feature assignments.

\section{Control experiments}

\subsection{Pure CH$_3$CHO ice deposition and subsequent H-atom addition}

\begin{figure}[t]
	\begin{center}
		\includegraphics[width=90mm]{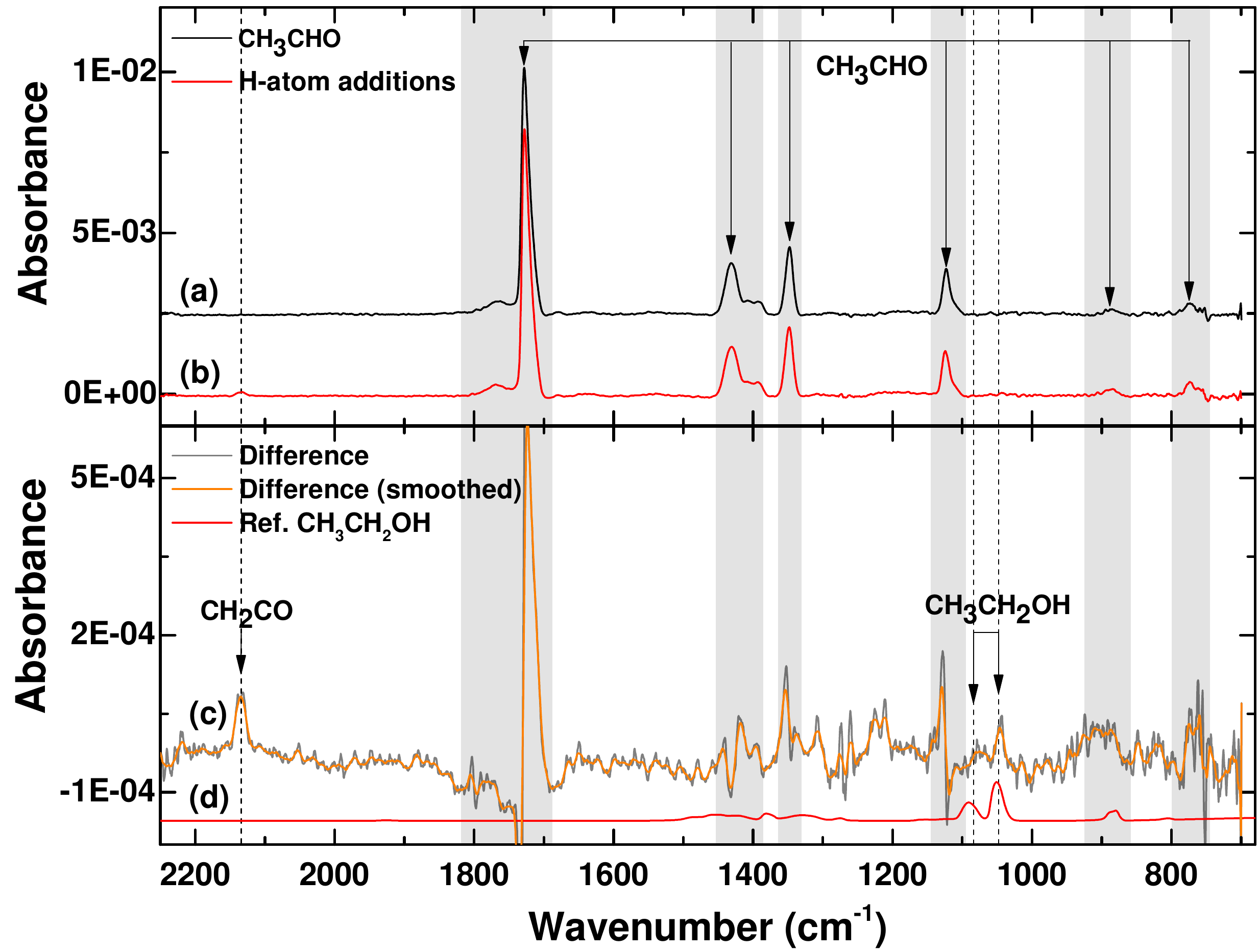}
		\caption{IR spectra of (a) pure CH$_3$CHO ice, (b) subsequent hydrogenation of predeposited ice with a H-atom flux of 1.2$\times$10$^{13}$ atoms cm$^{-2}$s$^{-1}$ for 30 min at 10 K, (c) RAIR difference spectrum obtained after hydrogenation of CH$_3$CHO ice, and (d) reference spectrum of CH$_3$CHO ice. The dashed lines and arrows are shown for the vibrational peaks of the studied molecules. IR spectra are offset for clarity}
		\label{Fig5}
	\end{center}
\end{figure}
Figure \ref{Fig5} shows the IR spectra obtained for (a) pure acetaldehyde ice deposited at 10 K and (b) subsequent hydrogenation of this predeposited ice, exposed to H-atoms with a flux of 1.2$\times$10$^{13}$ atoms cm$^{-2}$s$^{-1}$ for 30 min (exp. c1). IR features of (a) pure acetaldehyde ice has vibrational transitions at 1728, 1431, 1348, 1124, 885, and 774 cm$^{-1}$. These features are shifted compared to the previously shown results for acetaldehyde in C$_2$H$_2$ + O$_2$ + H (exp. 3) in Section 3.1, as the latter are recorded in a H$_2$O-rich environment. For example, the C=O stretching mode moves from 1728 to 1705 cm$^{-1}$ and the C-C stretching mode migrates oppositely from 1124 to 1132 cm$^{-1}$. A similar red (blue) shift for the C=O (C-C) bond of acetaldehyde in H$_2$O ices has been reported in \cite{TerwisschavanScheltinga2018} (Appendix B). These noticeable peak shifts as a function of a surrounding H$_2$O matrix can be used as a spectral tool to interpret future solid-state astronomical observations in mid-IR regions, in particular by the JWST. 

Besides these acetaldehyde IR peaks, there are no other features that can be assigned to vinyl alcohol implying that the efficiency of converting acetaldehyde to vinyl alcohol is negligible during ice deposition under our experimental conditions. In the IR spectrum (b), recorded after 30 min of H-atom exposure, the peak intensity of acetaldehyde decreases, and a few small bands located around 2135 and 1046 cm$^{-1}$ appear. A smoothed IR difference spectrum obtained after H-atom addition, represented by spectrum (c), allows one to distinguish the initial species depletion and new product formation, which are shown as negative and positive peaks, respectively. For example, ketene is firmly detected at 2135 cm$^{-1}$ and ethanol is shown at 1046 and 1085 cm$^{-1}$, which are the two most strong IR signatures of ethanol according to the reference spectrum (d). A tiny IR peak located around 886 cm$^{-1}$ may also originate from ethanol, but it is significantly overlapped with an acetaldehyde peak. 

\subsection{Simultaneous deposition of CH$_3$CHO + H and CH$_3$CHO + O$_2$ + H}

\begin{figure}[t]
	\begin{center}
		\includegraphics[width=90mm]{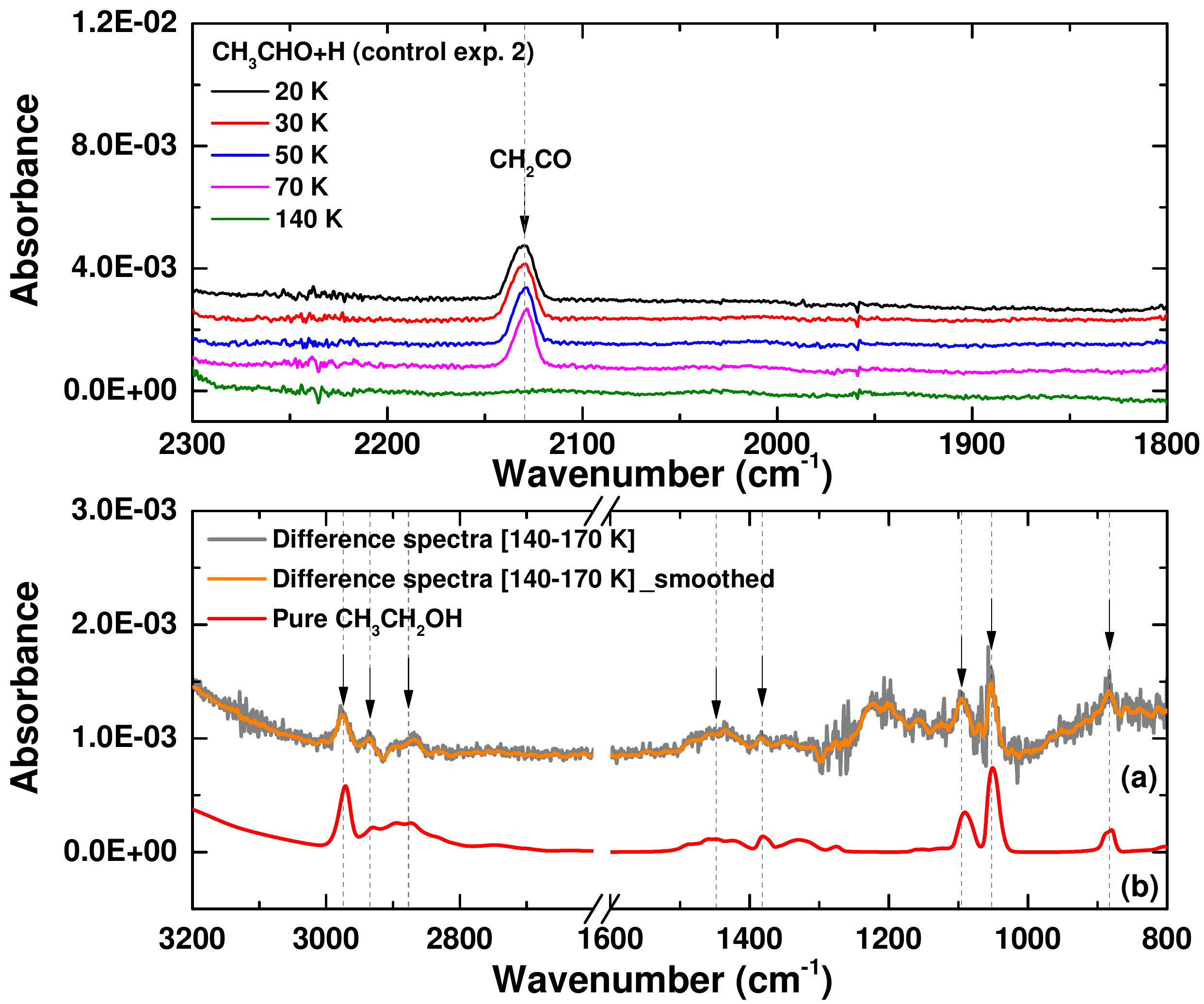}
		\caption{Upper: TPD-IR spectra obtained after the codeposition of CH$_3$CHO + H (exp. c2) at 10 K. The applied deposition ratio of CH$_3$CHO:H was 2:40, and the used H-atom flux was 1.2$\times$10$^{13}$ atoms cm$^{-2}$s$^{-1}$ Bottom: (a) TPD-IR difference spectrum for the range from 140 to 170 K obtained after the codeposition of CH$_3$CHO + H ice mixtures in a ratio of CH$_3$CHO:H=2:40 at 10 K. (b) Reference spectrum of CH$_3$CHO. The dashed lines and arrows are shown for the vibrational peaks of the studied molecules. IR spectra are offset for clarity.}
		\label{Fig7}
	\end{center}
\end{figure}

Figure \ref{Fig7} presents the IR spectrum of the acetaldehyde hydrogenation products (i.e., ketene and ethanol) during an ongoing TPD experiment (20-140 K) after 360 min of CH$_3$CHO + H (exp. c2) codeposition at 10 K. In the upper panel, it can be seen that the area of the ketene absorption peak located around 2130 cm$^{-1}$ remains constant up to 70 K and decreases at temperatures between 70 and 140 K. This is consistent with the desorption temperature range of 70-130 K for ketene as reported by \cite{Maity2014}. In the bottom panel, a smoothed IR difference spectrum (a) is shown, that was obtained between 140 and 170 K where the sublimation of CH$_3$CH$_2$OH ice is expected. It shows multiple spectral features at 2976, 2938, $\sim$2877 (broad), $\sim$1444 (broad), 1379, 1095, 1054, and 883 cm$^{-1}$. It is important to note that some of these peaks with higher absorption intensity are clearly detected in the IR spectrum obtained at 10 K. The resulting IR difference spectrum (between 140 and 170 K) excludes the spectral features originating from acetaldehyde, which is the most abundant species and desorbs between 90 and 130 K. The peak position and relative intensity in the IR difference spectrum all compare well with the reference spectrum (b) of pure ethanol. 
\begin{figure}[b]
	\begin{center}
		\includegraphics[width=90mm]{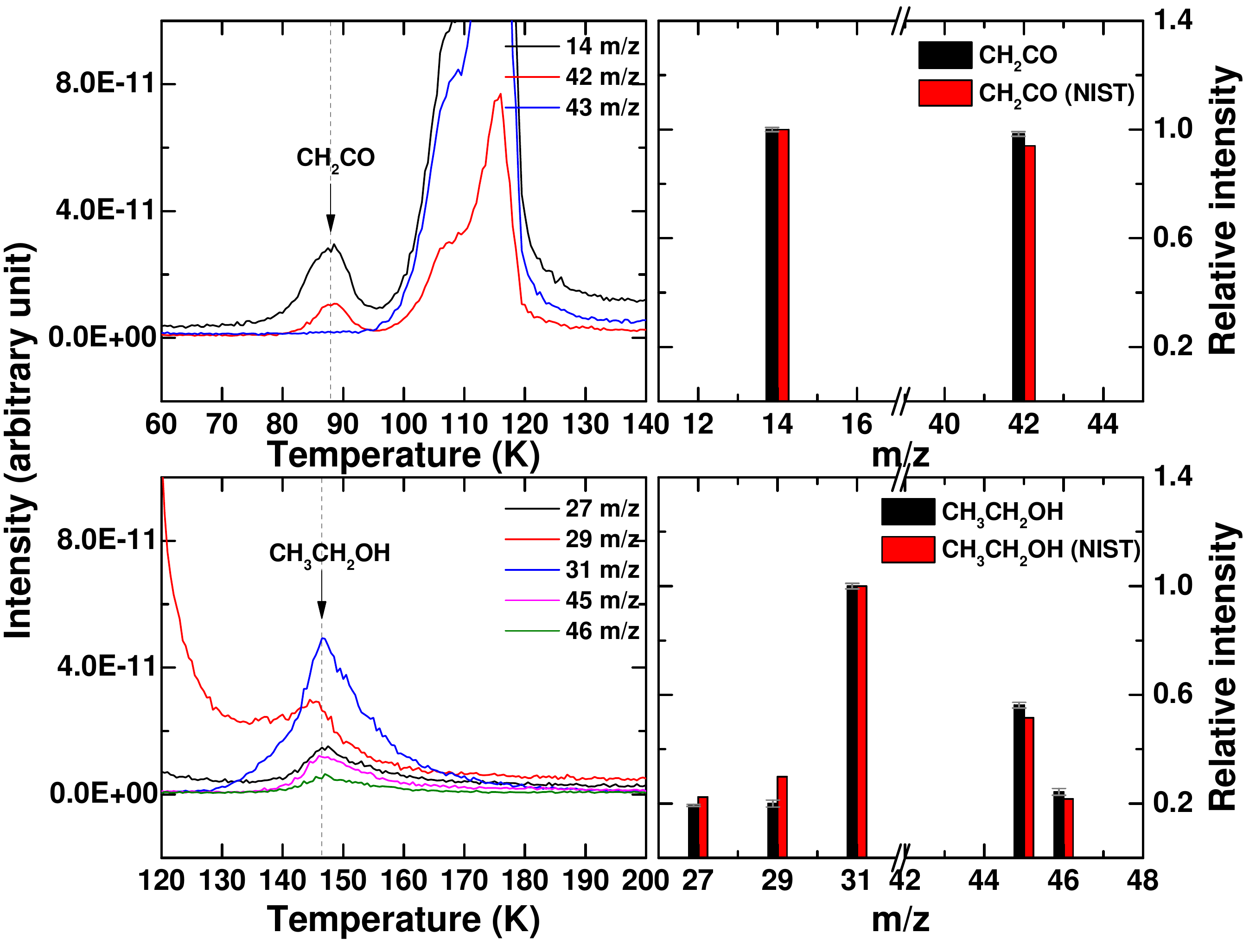}
		\caption{Left: TPD mass spectra obtained after 360 min of codeposition of CH$_3$CHO + H (exp. c2) at 10 K. The applied deposition ratio of CH$_3$CHO:H was 2:40, and the used H-atom flux was 1.2$\times$10$^{13}$ atoms cm$^{-2}$s$^{-1}$. Only relevant m/z channels are shown. The arrows and dashed lines indicate the peak position of the corresponding COMs. Right: comparison of the ionization fragmentation pattern of the detected product (black) with the standard pattern in the NIST database (red).}
		\label{Fig8}
	\end{center}
\end{figure}

Additional identifications of ketene and ethanol are independently realized by the QMS and shown in Fig. \ref{Fig8}. Two sets of mass signals are selected to represent the main product fragments of ketene (such as 14 and 42 m/z, upper panel) and of ethanol (such as 27, 29, 31, 45, and 46 m/z, bottom panel). Besides the parent species (acetaldehyde) desorbing peak at $\sim$110 K, both mass fragments of ketene (i.e., 14 and 42 m/z) start sublimating around 70 K and reach their peak at $\sim$88 K \citep{Maity2014}. The desorption profiles of ethanol fragments (i.e., 27, 29, 31, 45, and 46 m/z) are also found in the temperature range of 130-180 K and reach a maximum at 146 K. Both desorption temperatures are $\sim$10 K lower than the values observed in the C$_2$H$_2$ + O$_2$ + H experiment (exp. 3), in agreement with the fact that in H$_2$O(H$_2$O$_2$)-free ice mantles, the binding energies are lower. In the right-hand panels of Fig. \ref{Fig8}, the derived mass fragmentation patterns are shown that are comparable to the standard data of ketene (upper panel) and ethanol (bottom panel) as available from the NIST database. The mismatches (<10\% with respect to the maximum mass signals) are due to the baseline selection and high background signal from the residual gases in the chamber. As aforementioned, there is no proof for vinyl alcohol formation (i.e., there are no mass fragments for vinyl alcohol showing up at the corresponding desorption temperature, which is located between acetaldehyde and ethanol) in the QMS spectrum confirming the earlier IR analysis; conversion from acetaldehyde to vinyl alcohol in our experimental time-scale is very unlikely, even for the relatively high temperature during a TPD experiment. Both IR and QMS data obtained during the TPD experiment firmly support the identification of ketene and ethanol, and support the idea that these can be formed in space through the interaction of acetaldehyde with H-atoms on grain surfaces at 10 K.

\begin{figure}[t]
	\begin{center}
		\includegraphics[width=90mm]{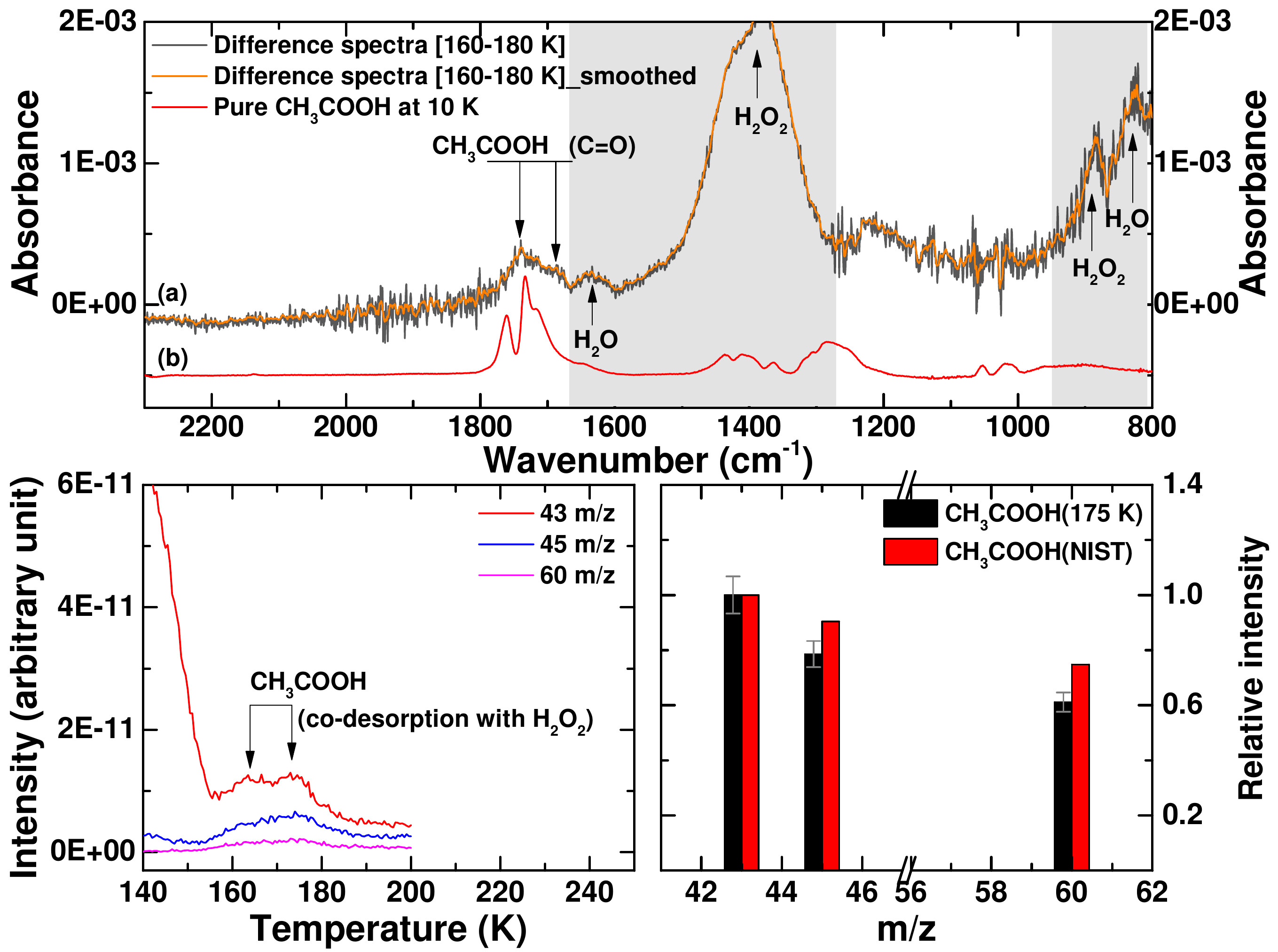}
		\caption{Upper: (a) smoothed TPD-IR difference spectrum for the range from 160 to 180 K obtained after the codeposition of CH$_3$CHO + O$_2$ + H (exp. c3) at 10 K. The applied deposition ratio of CH$_3$CHO:O$_2$:H was 2:10:40, and the used H-atom flux was 1.2$\times$10$^{13}$ atoms cm$^{-2}$s$^{-1}$ (b) reference IR spectrum of CH$_3$COOH. The IR spectra are offset for clarity. The shadow region indicates the codesorption peaks of H$_2$O$_2$ or H$_2$O.  Bottom: (Left) TPD mass spectra obtained after codeposition of CH$_3$CHO + O$_2$ + H ice mixtures in a ratio of 2:10:40 at 10 K. Only relevant m/z channels are shown. The arrows indicate the peak position of the acetic acid. (Right) A comparison of the ionization fragmentation pattern of the detected acetic acid (black) with the data in the NIST database (red).}
		\label{Fig9}
	\end{center}
\end{figure}

Figure \ref{Fig9} shows IR (upper panel) and QMS (bottom panels) spectra during TPD experiment after a 360-min codeposition experiment of CH$_3$CHO + O$_2$ + H (exp. c3) at 10 K. A smoothed IR difference spectrum obtained between 160 and 180 K implies the appearance of CH$_3$COOH by presenting the multiple C=O stretching modes in the range of 1780-1600 cm$^{-1}$ along with the codesorption peaks of H$_2$O$_2$ and trapped H$_2$O. These features have been suggested to originate from the acetic acid (monomer: 1770 and 1742 cm$^{-1}$ as well as dimer: 1716 and 1685 cm$^{-1}$) \citep{Bahr2006,Bertin2011}. These spectral features of acetic acid might hint at multiple sites where acetic acid is located or buried. An unambiguous spectral assignment of acetic acid is outside the scope of this work and we refer to detailed work by \cite{Bahr2006} and \cite{Bertin2011}. Rest peaks of acetic acid are overlapped or hidden in the codesorbing H$_2$O$_2$ and H$_2$O bands. However, the QMS data, typically with a higher detection sensitivity than RAIRS, shows CH$_3$COOH desorption in the same temperature range (i.e., 160-180 K) and the corresponding mass fragmentation pattern is generally consistent with the standard pattern in the NIST database in the bottom panels of Fig. \ref{Fig9}. The IR and QMS spectral results from the TPD experiment support the tentative detection of CH$_3$COOH in experiment C$_2$H$_2$ + O$_2$ + H (exp. 3). 

\end{appendix}
\end{document}